\documentclass[debug]{rmaa}
\usepackage{paralist}
\usepackage{changepage}
\usepackage{psfrag,color}
\usepackage{longtable}
\usepackage{float}
\usepackage[]{color}
\title{Identification of Planetary and Proto-Planetary Nebulae candidates through \emph{AKARI} infrared photometry}
\author{
R. A. Marquez-Lugo\altaffilmark{1,2} 
S.N. Kemp\altaffilmark{1,2} 
G. Ramos-Larios\altaffilmark{1,2} 
A. Nigoche-Netro\altaffilmark{1,2} 
S.G. Navarro\altaffilmark{1,2} 
L.J. Corral\altaffilmark{1,2}}
\altaffiltext{1}{Instituto de Astronomía y Meteorología, Universidad de Guadalajara}

\altaffiltext{2}{Departamento de Física, Universidad de Guadalajara}

\shortauthor{Marquez-Lugo et al.}
\shorttitle{RevMexAA Main Journal Demo Document}

\fulladdresses{

\item Marquez-Lugo: Instituto de Astronomía y Meteorología, Av. Vallarta 2602
Guadalajara, Jalisco, México. 44130.

(alejandro.marquez@academicos.udg.mx).}

\listofauthors{R.A. Marquez-Lugo, Kemp, S.N., Ramos-Larios, G., Nigoche-Netro, A., Navarro, S.G., & Corral, L.J.}
\indexauthor{Marquez-Lugo, R.A.}
\indexauthor{Kemp, S.N.}
\indexauthor{Ramos-Larios, G.}
\indexauthor{Nigoche-Netro, A.}
\indexauthor{Navarro, S.G.}
\indexauthor{Corral, L.J.}
\abstract{
We utilized photometric data from the space telescope \emph{AKARI} to identify potential planetary nebulae (PNe) and proto-planetary nebulae (PPNe) candidates. Using the colour-colour diagram, we found a region with a high concentration of established PNe and PPNe, comprising about 95\% of the objects. Based on this, we identified 67 objects within this region that lack definitive classification in existing literature, suggesting they are promising candidates. We conducted Spectral Energy Distribution (SED) analysis and morphological investigations using imagery from various observatories and satellites. Finally, we present a list of 65 potential PNe and PPNe candidates.
}

\resumen{Utilizamos datos fotométricos del telescopio espacial \emph{AKARI} para identificar posibles candidatos a nebulosas planetarias (PNe) y proto-nebulosas planetarias (PPNe). Utilizando el diagrama de color-color, encontramos una región con una alta concentración de PNe y PPNe establecidas, que comprende aproximadamente el 95\% de los objetos. Basados en esto, identificamos 67 objetos dentro de esta región que carecen de clasificación definitiva en la literatura existente, sugiriendo que son candidatos prometedores. Realizamos análisis de Distribución de Energía Espectral (SED) e investigaciones morfológicas utilizando imágenes de varios observatorios y satélites. Finalmente, presentamos una lista de 65 posibles candidatos a PNe y PPNe.
}

\addkeyword{(ISM:) planetary nebulae: general}
\addkeyword{stars: AGB and post-AGB}
\addkeyword{infrared: ISM}

\begin{document}
\maketitle
\section{Introduction}
\label{sec:intro}

The identification of extended planetary nebulae (PNe) has been carried out by morphological or spectral analysis \citep{1984ASSL..107.....P}.  However, in the case of unresolved sources, photometric methods, such as \emph{IRAS} (Infrared Astronomical Satellite) color-color diagrams, have been  employed {\citep{1988A&A...205..248P,1988A&AS...76..317P,1993IAUS..155...40V,1995A&A...299..238V,1989A&A...214..139M,1990A&AS...82..497G,2009A&A...501.1207R,2019MNRAS.488.3238A} or 2MASS (Two Micron All-Sky Survey) and \emph{WISE} (Wide-Field Infrared Survey Explorer) for the identification and classification of symbiotic stars \citep{2019MNRAS.483.5077A}. The available data from the infrared astronomical mission \emph{AKARI} \citep{2007PASJ...59S.369M} has been shown to be useful in the study of known PNe and proto-planetary nebulae (PPNe) \citep{2009ASPC..418..439C, 2009ASPC..418..137C, 2011AJ....141..111C, 2011RMxAA..47...83P, 2011EP&S...63.1051P, 2012PKAS...27..259O, 2016AJ....151...93O, 2019IAUS..343..522U}, which is why we have used photometric data from the \emph{AKARI} Point Source Catalog (\emph{AKARI} PSC) to identify new PNe and PPNe candidates.

\begin{figure}[H]
\begin{changemargin}{-1cm}{-1cm}
\includegraphics[scale=0.55, trim=0 0 0 0,clip]{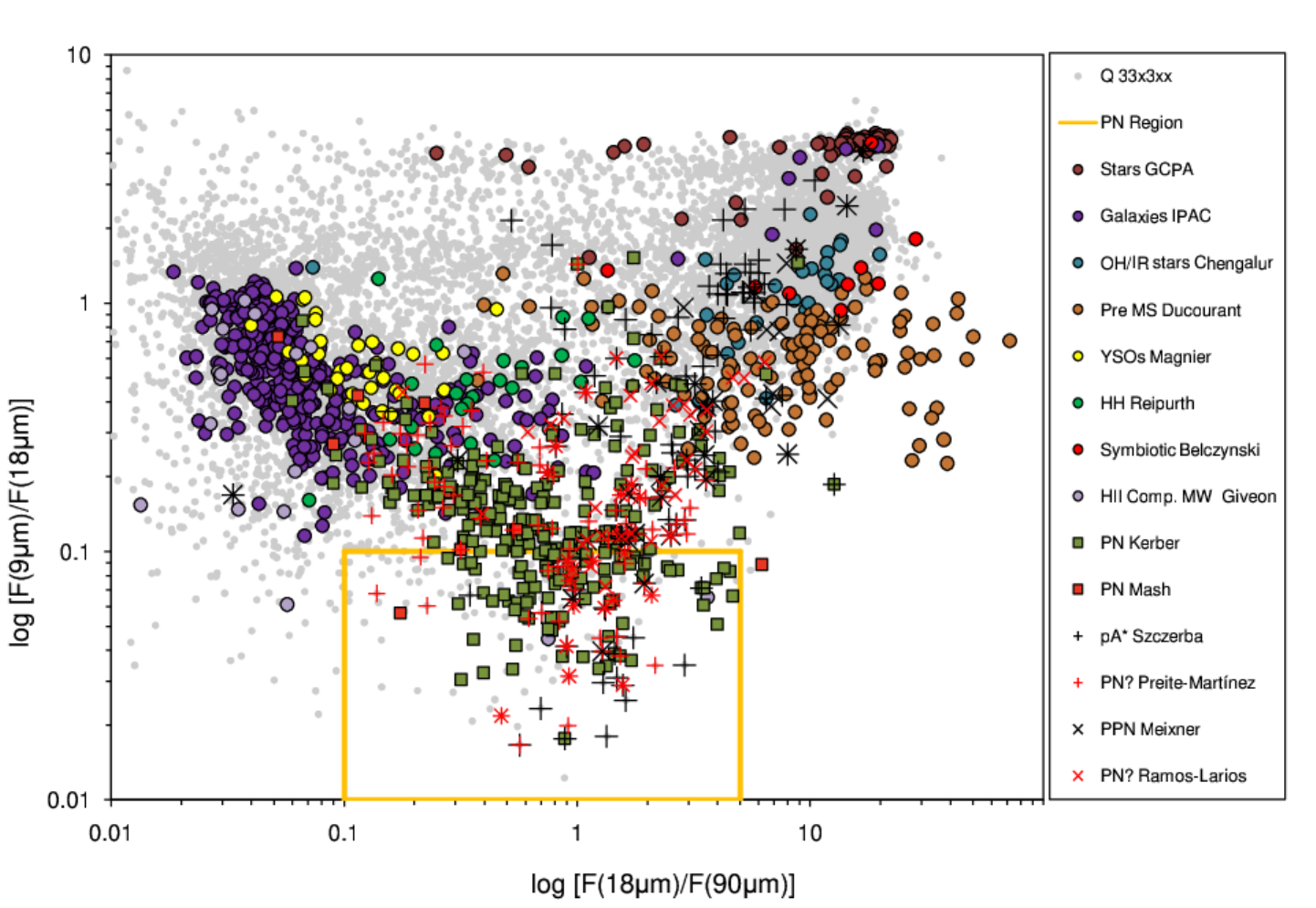}
\end{changemargin}
\caption{ \emph{AKARI} PSC color-color diagram. The \emph{AKARI} PSC sources with Q33x3xx are shown as grey points, and the right hand panel indicates other, superimposed symbols (see text for details). The area delineated in yellow contains a concentration of PNe, PPNe, post-AGB stars, and candidates for all these categories, and excludes an enormous proportion of other types of astronomical objects.}
\label{Colour-Colour Diagram}
\end{figure} 

We utilized widely recognized catalogues encompassing Planetary Nebulae (PNe), galaxies, Young Stellar Objects (YSOs), main sequence and pre-main sequence stars, symbiotic stars, OH/IR stars, Herbig-Haro objects (HH), and HII compact regions. The objective was to pinpoint the positions of these sources within an \emph{AKARI} Point Source Catalogue (PSC) colour-colour diagram. Our focus was on identifying a distinct region where Planetary Nebulae (PNe) are prominently featured and relatively segregated from other categories of astronomical entities.

\section{Observations}
\label{obs}
The \emph{AKARI} PSC photometry has been accessible to the public since March 2010. In addition, mission image data has been released, starting with the \emph{AKARI} far-infrared all-sky survey maps, available since December 2014, and subsequently the \emph{AKARI}/IRC Pointed Observation Images from March 2015 (IRC, InfraRed Camera). However, owing to the suboptimal resolution of \emph{AKARI} images for the present study, we employed image data from other infrared and optical projects such as the \emph{Spitzer Space Telescope}, The INT Photometric H $\alpha$ Survey of the Northern Galactic Plane (IPHAS), 2MASS, and \emph{WISE}. These alternative datasets were utilized to examine the spatial emission profiles of PNe candidates and, as a preliminary step, to analyze their morphology (for detailed information, refer to Section~\ref{mor}).

To generate the color-color diagram for distinguishing between various types of astronomical objects, we employed photometry data from the PSC of the \emph{AKARI} space telescope (previously known as ASTRO-F or IRIS Infra Red Imaging Surveyor, launched on February 21, 2006). This dataset comprises two components: one obtained from the IRC featuring two bands centered at 9$\mu$m and 18$\mu$m \citep{2010A&A...514A...1I}, and the other from the Far-Infrared Surveyor (FIS) with four bands centered at 65$\mu$m, 90$\mu$m, 140$\mu$m, and 160$\mu$m \citep{2007PASJ...59S.389K}.

For constructing the Spectral Energy Distributions (SEDs) we utilized data not only from \emph{AKARI}, but also from: US Naval Observatory B catalogue (USNO-B), compiled from the digitalisation of various photographic sky surveys plates by the Precision Measuring Machine (PMM), located at the US Naval Observatory Flagstaff Station (NOFS), from 350 nm to 900 nm between 1949 and 2002 \citep{2003AJ....125..984M}; 2MASS, with three bands J (1.25$\mu$m), H (1.65$\mu$m) and K (2.17$\mu$m) from the 1.3 m telescopes at Mt. Hopkins and Cerro Tololo Inter-American Observatory (CTIO) Chile \citep{1992ASPC...34..203K}; the Deep Near Infrared Southern Sky Survey (DENIS) \citep{2000A&AS..141..313F} with three bands i (0.802$\mu$m), J (1.248$\mu$m) and $K_s$ (2.152$\mu$m); the Midcourse Space Experiment (MSX) \citep{1996AJ....112.2862E}, from the telescope SPIRIT III aboard the Ballistic Missile Defense Organization (BMDO) with the bands A (8.28$\mu$m), C (12.13$\mu$m), D (14.65$\mu$m) and E (21.3$\mu$m) and the Infrared Array Camera (IRAC) (Fazio et al. 2004) aboard the \emph{Spitzer Space Telescope} with the bands I1 (3.6$\mu$m), I2 (4.5$\mu$m), I3 (5.8$\mu$m) and I4 (8.0$\mu$m).

The resolution of the images of the \emph{AKARI} far-infrared all-sky survey maps, with pixel sizes of the detectors of 26.8$\arcsec$$\times$26.8$\arcsec$ for the short wavelength bands N60 and WIDE-S, and 44.2$\arcsec$$\times$44.2$\arcsec$ for the long wavelength bands N160 and WIDE-L \citep{2015PASJ...67...50D} is too poor to study the morphology and vicinity of the candidates, therefore \emph{WISE} (Wide-Field Infrared Survey Explorer, \citet{2010AJ....140.1868W}) images were used for this purpose. Furthermore, radial emission profiles of sources were derived from \emph{Spitzer} IRAC, 2MASS and IPHAS data.

\section{Searching for candidates}

\subsection{The \emph{AKARI} Point Source Catalogue}

The \emph{AKARI} PSC comprises two components: the \emph{AKARI} FIS Bright Source Catalogue \citep{2010yCat.2298....0Y} and the \emph{AKARI} IRC Point Source Catalogue \citep{2010yCat.2297....0I}. The AKARI IRC Catalogue encompasses data from two bands centered at 9$\mu$m and 18$\mu$m. The 9$\mu$m band includes information for 844,649 sources, while the 18$\mu$m band comprises data for 194,551 sources In total, there is information for 870,973 unique sources, with 168,227 sources appearing in both bands. On the other hand, the \emph{AKARI} FIS Catalogue provides photometric data for 427,071 sources, each observed in at least one of the four bands centered at 65$\mu$m, 90$\mu$m, 140$\mu$m, and 160$\mu$m. Among these, the 90$\mu$m band, with data for 373,553 sources, emerges as the most crucial for subsequent analyses due to its heightened sensitivity.

In summary the \emph{AKARI} PSC provides photometric information for six bands. Each data comes with a quality indicator called Qualityflag (FQUAL). This indicator can take values from 0 to 3. FQUAL=3 is the highest quality, indicating a confirmed source. FQUAL=2 indicates a confirmed source with problems, a source with a very low  flux level or a possible false detection, for example a ``side-lobe'' effect. FQUAL=1 implies that the source was not confirmed or that the flux value is unreliable. Finally FQUAL=0 is a non-detection. For our inquiry, we exclusively considered sources with a Qualityflag of FQUAL=3 in the 9$\mu$m, 18$\mu$m, and 90$\mu$m bands, and any Qualityflag in the 65$\mu$m, 140$\mu$m, and 160$\mu$m bands. We denote this specific combination of Qualityflag as Q33x3xx, representing the bands 9$\mu$m, 18$\mu$m, 65$\mu$m, 90$\mu$m, 140$\mu$m, and 160$\mu$m, respectively. This selection resulted in a total of 9,900 sources.

The \emph{AKARI} photometric information was obtained from DARTS/Akari at ISAS/JAXA  http://darts.jaxa.jp/ir/akari/cas/tools/search/crossid.html 
 using  \emph{AKARI}-CAS \citep{2011PASP..123..852Y}.

\subsection{Catalogs for source discrimination}

\label{catlogs}

We have cross-referenced these 9,900 sources from the \emph{AKARI} PSC, which report high-quality photometry, against catalogues of various types of astrophysical objects from VizieR and some bibliographic references to establish correspondences. The catalogues employed for this purpose were as follows: 

General catalogue of stars with the system of photoelectric astrolabes (Working Group Of GCPA 1992), Catalogued galaxies and quasars observed in the \emph{IRAS} survey, Version 2 \citep{1989IRASG....c...0F}, New OH/IR stars from color-selected \emph{IRAS} sources. 3: A complete survey  \citep{1996yCat..20890189C}, Pre-main sequence star Proper Motion Catalogue \citep{2005yCat..34380769D}, Transitional YSOs (Young Stellar Objects) candidates from flat-spectrum \emph{IRAS} sources \citep{2000yCat..33520228M}, A General Catalogue of Herbig-Haro Objects, 2nd Edition \citep{2000yCat.5104....0R}, A catalogue of symbiotic stars \citep{2001yCat..41460407B}, A New Catalogue of Radio Compact H II Regions in the Milky Way. II. The 1.4 GHz Data \citep{2005AJ....129..348G}, Galactic Planetary Nebulae and their central stars. I. An accurate and homogeneous set of coordinates \citep{2004yCat..34081029K}, MASH  (Macquarie/AAO/Strasbourg $H_\alpha$)  Catalogues of Planetary Nebulae \citep{2006yCat.5127....0P}, Torun catalogue of post-AGB (post-Asymtotic Giant Branch) and related objects \citep{2007yCat..34690799S}, A Mid-Infrared Imaging Survey of Proto-Planetary Nebula Candidates \citep{1999ApJS..122..221M}, Possible new planetary nebulae in the \emph{IRAS} Point Source Catalogue  \citep{1988A&AS...76..317P} and \emph{IRAS} identification of Post-AGB and PNe candidates \citep{2009A&A...501.1207R}. The selection was carried out with the aim of encompassing a broad variety of astronomical objects, facilitating a comprehensive comparison of their infrared emissions and positions in colour-colour maps, in relation to the emission and position of PNe and PPNe.

\subsection{Color--color diagram}

After reviewing the work of \citet{1990A&AS...82..497G}, taking into account the available AKARI bands, and thoroughly exploring potential color combinations, we have chosen the following color definitions.:

$$
x= log{F(18\mu m)\over
     F(90\mu m)},
$$

and

$$
y= log{F(9\mu m)\over
     F(18\mu m)}.
$$ 
\hspace{1cm}

In Figure \ref{Colour-Colour Diagram}, we have plotted the values for log[F(18$\mu$m)/F(90$\mu$m)] against log[F(9$\mu$m)/F(18$\mu$m)] for the 9,900 sources in the \emph{AKARI} PSC with photometry quality Q33x3xx.  Superimposed symbols represent objects also present in the catalogues mentioned above (Subsection~\ref{catlogs}). The diagram delineates distinct areas corresponding to different types of objects. Notably, there is a defined region (marked by the yellow line) where known PNe and PNe candidates are concentrated. This region, referred to as the PNe region, is situated between:

$$
0.1 \leq log{F(18\mu m)\over
     F(90\mu m)} \leq 5,
$$

and

$$
0.01 \leq log{F(9\mu m)\over
     F(18\mu m)} \leq 0.1.
$$ 

In addition to the objects from the catalogues of known PNe and PNe candidates, only two objects from the `New Catalog of Radio Compact H II Regions in the Milky Way' are located in the PNe region. No objects from any of the catalogues representing other types were identified in this region. The PNe region includes 215 sources from the \emph{AKARI} PSC with Q33x3xx. Excluding the previously identified PNe, there are 67 sources that can in principle be regarded as potential PNe candidates (Table~1).

The selection of these limits is conservative. Confirming the proposed candidates in this study could warrant an extension of the PN region, particularly by increasing the value on the ordinate axis to 0.2 to encompass more potential candidates. However, this adjustment might introduce a risk of potential contamination of the PN region with galaxies.

\begin{changemargin}{-2cm}{0cm}
\setlength\LTleft{-3cm}
\begin{small}
\begin{longtable}{lllcclllll}
\caption{\emph{AKARI} Planetary Nebulae Candidates.}\\
\hline 
No & IRC & G.C. & RA (2000) & DEC (2000) & IDENTIFICATION & DIST & TYPE \\ 
 &  &  & H M S & D M S &  & arcsec &  \\ 
\hline
\endfirsthead
\multicolumn{8}{c}
{TABLE 1 \ -- \textit{AKARI \emph{Planetary Nebulae Candidates} (Continued from previous page)}}\\
\hline 
No & IRC & G.C. & RA (2000) & DEC (2000) & IDENTIFICATION & DIST & TYPE \\ 
 &  &  & H M S & D M S &  & arcsec &  \\ 
\hline
\endhead
\hline \multicolumn{8}{r}{\textit{Continued on next page}}\\
\endfoot
\hline
\endlastfoot
\hline
1	&	200475060	&	G001.30+04.07	&	17 33 13.13	&	 -25 40 17.7	&	-	&	-	&	-	\\
2	&	200557576	&	G006.43-01.92	&	18 07 31.86	&	 -24 20 21.1	&	-	&	-	&	-	\\
3	&	200455657	&	G008.55+11.49	&	17 23 11.65	&	 -15 37 15.9	&	[M81] I-556 	&	2.36	&	Em* 	\\
	&		&		&		&		&	\emph{IRAS} 17203-1534 	&	3.85	&	pA* 	\\
4	&	200535439	&	G008.85+01.69	&	17 59 05.08	&	 -20 27 24.4	&	SCHB 258 	&	0.61	&	Mas 	\\
5	&	200556938	&	G009.10-00.39	&	18 07 20.97	&	 -21 16 11.2	&	SCHB 292 	&	0.37	&	Mas 	\\
	&		&		&		&		&	\emph{IRAS} 18043-2116 	&	3.49	&	IR 	\\
6	&	200562588	&	G011.41+00.43	&	18 09 04.73	&	 -18 50 42.6	&	\emph{IRAS} 18061-1851 	&	4.4	&	IR 	\\
7	&	200585006	&	G014.03-00.66	&	18 18 22.46	&	 -17 04 00.6	&	-	&	-	&	-	\\
8	&	200580061	&	G015.70+00.77	&	18 16 25.80	&	 -14 55 13.0	&	-	&	-	&	-	\\
9	&	200605366	&	G017.01-01.23	&	18 26 15.79	&	 -14 42 26.5	&	OH 17.0 -1.2 	&	2.46	&	Mas 	\\
	&		&		&		&		&	[TVH89] 301 	&	2.46	&	Mas 	\\
	&		&		&		&		&	PN PM 1-232 	&	5.17	&	PN? 	\\
10	&	200600445	&	G020.41+01.11	&	18 24 18.70	&	 -10 36 30.2	&	\emph{IRAS} 18215-1038 	&	2.76	&	IR 	\\
11	&	200627833	&	G020.51-02.07	&	18 35 59.44	&	 -11 59 39.1	&	-	&	-	&	-	\\
12	&	200628891	&	G025.26+00.25	&	18 36 29.20	&	 -06 42 46.7	&	GSC 05124-02611 	&	1.18	&	* 	\\
13	&	200637610	&	G027.54+00.41	&	18 40 06.57	&	 -04 36 41.9	&	-	&	-	&	-	\\
14	&	200643853	&	G028.34+00.12	&	18 42 37.09	&	-04 02 04.7	&	-	&	-	&	-	\\
15	&	200686203	&	G030.09-04.74	&	19 03 09.52	&	-04 41 08.4	&	-	&	-	&	-	\\
16	&	200675903	&	G036.07-00.17	&	18 57 46.14	&	 02 43 09.3	&	-	&	-	&	-	\\
17	&	200700351	&	G036.41-03.68	&	19 10 54.42	&	 01 24 41.6	&	2MASS J19105453+0124450 	&	0.5	&	Pe* 	\\
18	&	200689042	&	G036.43-01.91	&	19 04 38.65	&	 02 14 23.7	&	PK 036-01 3 	&	2.4	&	PN? 	\\
19	&	200680582	&	G037.28-00.23	&	19 00 10.88	&	 03 45 48.7	&	[WBH2005] G037.278-0.226 	&	1.27	&	Rad 	\\
	&		&		&		&		&	V* V1672 Aql 	&	1.61	&	V* 	\\
	&		&		&		&		&	2MASS J19001104+0345511 	&	3.51	&	IR 	\\
20	&	200703532	&	G038.56-03.11	&	19 12 47.88	&	 03 34 40.4	&	-	&	-	&	-	\\
21	&	200704956	&	G042.90-01.08	&	19 13 37.89	&	 08 21 47.9	&	-	&	-	&	-	\\
22	&	200697923	&	G043.03+00.14	&	19 09 30.06	&	 09 02 25.9	&	GPSR 043.028+0.140 	&	0.75	&	Rad 	\\
	&		&		&		&		&	MSX6C G043.0281+00.1406 	&	1.11	&	PN?	\\
23	&	200754442	&	G052.24-06.42	&	19 51 00.78	&	 13 58 20.4	&	V* TW Aql 	&	5.44	&	sr* 	\\
24	&	200732448	&	G057.11+01.46 	&	19 32 13.50	&	 22 04 57.6	&	2MASS J19321348+2204566  	&	0.99	&	Y*? 	\\
25	&	200734399	&	G057.81+01.46	&	19 33 41.50	&	 22 42 02.8	&	PN PM 1-306 	&	5.34	&	PN? 	\\
26	&	200765758	&	G068.20+00.24	&	20 01 59.87	&	 31 03 09.9	&	\emph{IRAS} 20000+3054 	&	4.8	&	IR 	\\
27	&	200775563	&	G072.25+00.26	&	20 12 17.32	&	 34 28 10.6	&	-	&	-	&	-	\\
28	&	200500801	&	G077.13+30.87	&	17 44 55.03	&	 50 02 40.2	&	V* V814 Her 	&	4.25	&	sr* 	\\
29	&	200789311	&	G077.92+00.22	&	20 28 30.55	&	 39 07 02.3	&	OH 2026+38 	&	0.96	&	Rad 	\\
30	&	200796994	&	G080.94+00.02	&	20 38 49.61	&	 41 25 22.0	&	-	&	-	&	-	\\
31	&	200778777	&	G083.19+06.64	&	20 15 57.33	&	 47 05 34.5	&	-	&	-	&	-	\\
32	&	200808628	&	G083.61-02.20	&	20 57 03.78	&	 42 06 26.6	&	-	&	-	&	-	\\
33	&	200828078	&	G096.54+01.36	&	21 35 43.92	&	53 53 12.4	&	-	&	-	&	-	\\
34	&	200862609	&	G111.71-02.13	&	23 23 13.49	&	 58 48 05.0	&	[PF2007] R3 	&	5.04	&	reg 	\\
35	&	200866080	&	G113.89-01.60	&	23 38 13.11	&	 59 58 42.7	&	-	&	-	&	-	\\
36	&	200865230	&	G115.21+04.32	&	23 34 21.97	&	 66 01 56.5	&	\emph{IRAS} 23321+6545 	&	5.72	&	C* 	\\
37	&	200033305	&	G135.28+02.80	&	02 43 28.32	&	 62 57 05.5	&	-	&	-	&	-	\\
38	&	200044584	&	G140.48+06.04	&	03 37 28.74	&	 63 06 27.2	&	PN PM 1-8 	&	2.44	&	PN? 	\\
	&		&		&		&		&	MSX6C G200.0789-01.6323 	&	5.19	&	Y*O 	\\
39	&	200184830	&	G282.84-01.25	&	10 11 16.59	&	 -57 44 10.6	&	-	&	-	&	-	\\
40	&	200194383	&	G286.32-00.70	&	10 36 01.61	&	 -59 07 44.6	&	\emph{IRAS} 10341-5852 	&	3.93	&	IR 	\\
	&		&		&		&		&	HD 303122 	&	4.76	&	* 	\\
41	&	200193271	&	G286.60-01.92	&	10 33 05.11	&	 -60 19 48.9	&	HD 305298 	&	2.6	&	* 	\\
42	&	200200079	&	G289.36-02.65	&	10 50 20.48	&	 -62 17 09.5	&	SPH 100 	&	4.12	&	EmO 	\\
	&		&		&		&		&	\emph{IRAS} 10484-6201 	&	4.38	&	IR 	\\
43	&	200212827	&	G292.89-01.20	&	11 23 11.95	&	 -62 20 25.8	&	-	&	-	&	-	\\
44	&	200215868	&	G293.45+00.07	&	11 31 08.80	&	 -61 19 07.0	&	-	&	-	&	-	\\
45	&	200212510	&	G294.04-04.71	&	11 22 18.58	&	 -66 01 46.8	&	\emph{IRAS} 11201-6545 	&	4.35	&	pA? 	\\
46	&	200235503	&	G299.54-00.02	&	12 21 45.12	&	 -62 41 50.1	&	\emph{IRAS} 12190-6225 	&	2.71	&	IR 	\\
47	&	200246253	&	G302.55-02.74	&	12 47 46.78	&	 -65 36 41.3	&	PN PM 1-69 	&	2.43	&	PN? 	\\
48	&	200254043	&	G304.57+00.72	&	13 05 27.82	&	 -62 06 37.4	&	-	&	-	&	-	\\
49	&	200259071	&	G305.79+00.28	&	13 16 11.59	&	 -62 27 14.2	&	\emph{IRAS} 13129-6211 	&	1.88	&	IR 	\\
50	&	200279625	&	G311.02+02.03	&	13 56 24.00	&	 -59 48 59.3	&	\emph{IRAS} 13529-5934 	&	4.96	&	pA? 	\\
51	&	200285648	&	G311.68-00.63	&	14 07 35.16	&	 -62 11 47.3	&	-	&	-	&	-	\\
52	&	200302084	&	G313.87-04.08	&	14 36 34.50	&	 -64 41 31.5	&	\emph{IRAS} 14325-6428 	&	0.89	&	pA? 	\\
53	&	200314558	&	G316.94-02.88	&	14 56 46.81	&	 -62 17 29.9	&	-	&	-	&	-	\\
54	&	200323862	&	G320.68+00.25	&	15 10 43.74	&	 -57 44 47.7	&	\emph{IRAS} 15068-5733 	&	3.6	&	IR 	\\
55	&	200327862	&	G321.02-00.70	&	15 16 41.23	&	 -58 22 28.6	&	\emph{IRAS} 15127-5811 	&	3.2	&	IR 	\\
56	&	200354340	&	G325.16-03.01	&	15 52 19.68	&	 -57 50 52.5	&	\emph{IRAS} 15482-5741 	&	4.31	&	pA? 	\\
57	&	200366578	&	G331.16+00.78	&	16 06 40.62	&	 -51 03 56.4	&	OH 331.16 +0.78 	&	1.35	&	Mas 	\\
58	&	200365730	&	G332.29+02.28	&	16 05 41.23	&	 -49 11 33.8	&	Caswell OH 332.295+02.280 	&	5.79	&	Mas 	\\
	&		&		&		&		&	Caswell CH3OH 332.295+02.280 	&	5.98	&	Mas 	\\
59	&	200392246	&	G335.06-01.16	&	16 32 39.87	&	 -49 42 14.0	&	EM* VRMF 55 	&	0.78	&	Em* 	\\
	&		&		&		&		&	IGR J16327-4940 	&	0.78	&	HXB 	\\
	&		&		&		&		&	\emph{IRAS} 16288-4935 	&	4.69	&	IR 	\\
60	&	200397316	&	G335.95-01.36	&	16 37 14.40	&	 -49 11 19.0	&	PN G335.9-01.3 	&	3.67	&	PN? 	\\
61	&	200389424	&	G336.24+00.51	&	16 30 10.70	&	 -47 42 37.8	&	-	&	-	&	-	\\
62	&	200412657	&	G340.43-00.37	&	16 50 08.91	&	 -45 09 26.4	&	-	&	-	&	-	\\
63	&	200420643	&	G342.44-00.23	&	16 56 39.52	&	 -43 30 49.6	&	-	&	-	&	-	\\
64	&	200413826	&	G350.43+07.61	&	16 51 06.42	&	 -32 23 00.5	&	-	&	-	&	-	\\
65	&	200422895	&	G352.57+07.30	&	16 58 27.11	&	 -30 55 06.3	&	2MASS J16582725-3055108 	&	4.98	&	IR 	\\
66	&	200489617	&	G355.61-01.16	&	17 39 27.29	&	 -33 16 41.2	&	-	&	-	&	-	\\
67	&	200571327	&	G359.81-06.87	&	18 12 58.61	&	 -32 30 04.1	&	-	&	-	&	-	\\
 \hline
 \end{longtable}
\end{small}
\end{changemargin}

\section{SIMBAD sources}

We referred to the SIMBAD database to gather information about our 67 potential candidates, as presented in Table~1. The search radius was set at 6 arcsec. 

Out of the 67 potential candidates, 30 did not yield any identification, while in 37 cases, at least one identification was obtained. Table~1 provides details such as the IRC identifier number, galactic coordinates, right ascension, declination, the SIMBAD identification, distance between the IRC source and the SIMBAD identification, and the concise SIMBAD object type (http://simbad.u-strasbg.fr/simbad/sim-display?data=otypes).

The total number of identifications is 50 (there may be more than one per object), of which 13 are infrared sources (IR), 7 maser sources (Mas), 7 possible PNe  (PN?), 4 post-AGB candidates (pA?), 3 stars (*), 3 radio sources (Rad), 2 emission-line stars (Em*), 2 semi-regular pulsating stars (sr*) and 1 identification  of each of the following types of objects: post-AGB-PPN (pA*), peculiar star (Pe*), variable star (V*), young stellar object candidate (Y*?), `region defined in the sky' (reg),  carbon star (C*), young stellar object (YSO), emission object (EmO) and high mass X-ray binary (HXB).

\begin{figure}[H]
\begin{changemargin}{-1cm}{-1cm}
\includegraphics[scale=0.37, trim=0 0 0 0,clip]{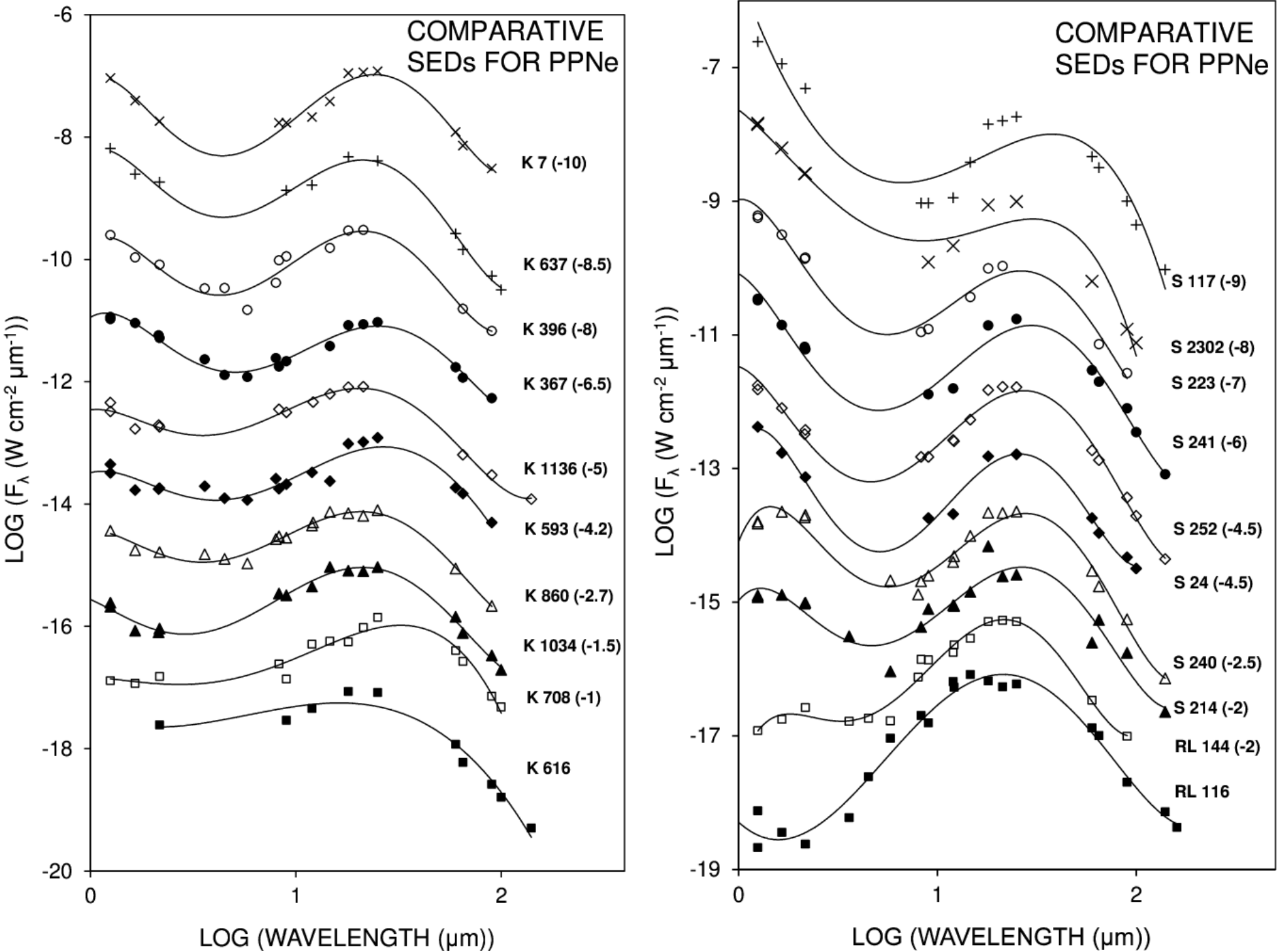}
\end{changemargin}
\caption{SEDs of established PPNe. The double-component structure is clearly seen in the SEDs. Each PPN is identified by the initials of the authors of the corresponding catalogues: Kerber, Szczerba, and Ramos-Larios, followed by the corresponding identification number. In order to avoid overlaps between different SEDs, an offset has been applied, the value of the offset is given in brackets after the PPN name.}
\label{Known PPN}
\end{figure}

As a PN is an ionized circumstellar shell around a hot compact star,  in the process of evolving between the AGB and white dwarf (WD) phase, it can be described in various ways depending on the particular characteristic under study.
For example, it may be described as a transition object (pA* or pA?) while the central star is ionizing the envelope, the spectrum of which shows strong emission lines (Em*) including those of ionized hydrogen (H II), the material expelled during the AGB mass-loss phase is heated by radiation or shocks, and therefore is an IR source (IR), and this material can contain large quantities of carbon, product of the triple-$\alpha$ reaction (C*). Besides, the central stars of PNe, which are terminating their AGB phase, can have brightness variations (sr* and V*), also the PPNe can show maser emission (Mas) due to the presence of water \citep{2019IAUS..343..527U}, and finally some rotational molecular transitions emit at radio wavelengths (Rad). So all these classifications do not rule out the possibility of the objects being PN or PPN candidates, therefore objects thus identified are not excluded from the potential candidate list.

On the other hand, classifications such as star (*), peculiar star (Pe*), region defined in the sky (reg) and emission object (EmO) are poorly defined and are not sufficient reasons to exclude sources  with such classifications from the potential candidate list.

\begin{figure}[H]
\begin{changemargin}{2cm}{-1cm}
\includegraphics[scale=0.5, trim=-0 0 0 0,clip]{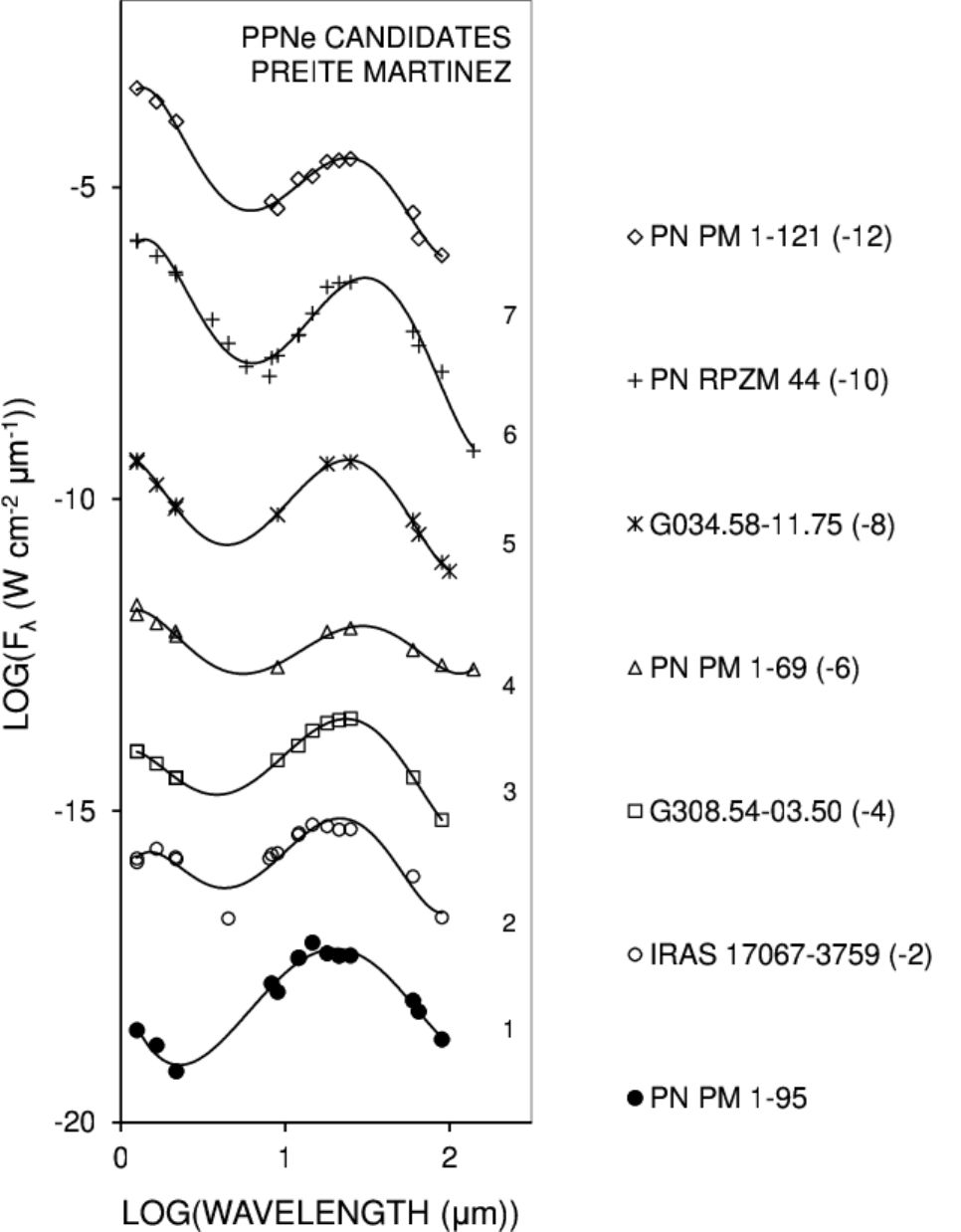}
\end{changemargin}
\caption{SEDs of PPN candidates from \citep{1988A&AS...76..317P}. The double-component structure is also clear in these candidates. In order to avoid overlaps between different spectra, an offset has been applied to the spectra. The value of the offset is given in brackets after the PPN name.}
\label{PM PPN}
\end{figure}

There are only three additional classifications corresponding to G057.11 +01.46, G140.48 +06.04 and G335.06 --01.16 (numbers 24, 38, and 59, respectively, in Table~1). The first source is designated as a candidate young stellar object (Y*?), but due to its unconfirmed status, we do not exclude it. The second source carries an additional classification as a PN candidate (PN?), which is why it remains in the list. The third is identified as a high-mass X-ray binary (HXB) and is also classified as an emission-line star (Em*) and an IR source (IR), the last two being consistent with PN candidacy.

Utilizing information from SIMBAD, there is no justification to exclude any of the 67 sources from the list of potential PNe candidates. Following this we present a thorough analysis of their Spectral Energy Distributions (SEDs) to discern their posible evolutionary state, complemented by a morphological examination of the sources and an assessment of their environment using archival images.

\begin{figure}[H] 
\begin{changemargin}{-2cm}{-2cm}
\centering
\includegraphics[scale=0.55, trim=0 0 0 0,clip]{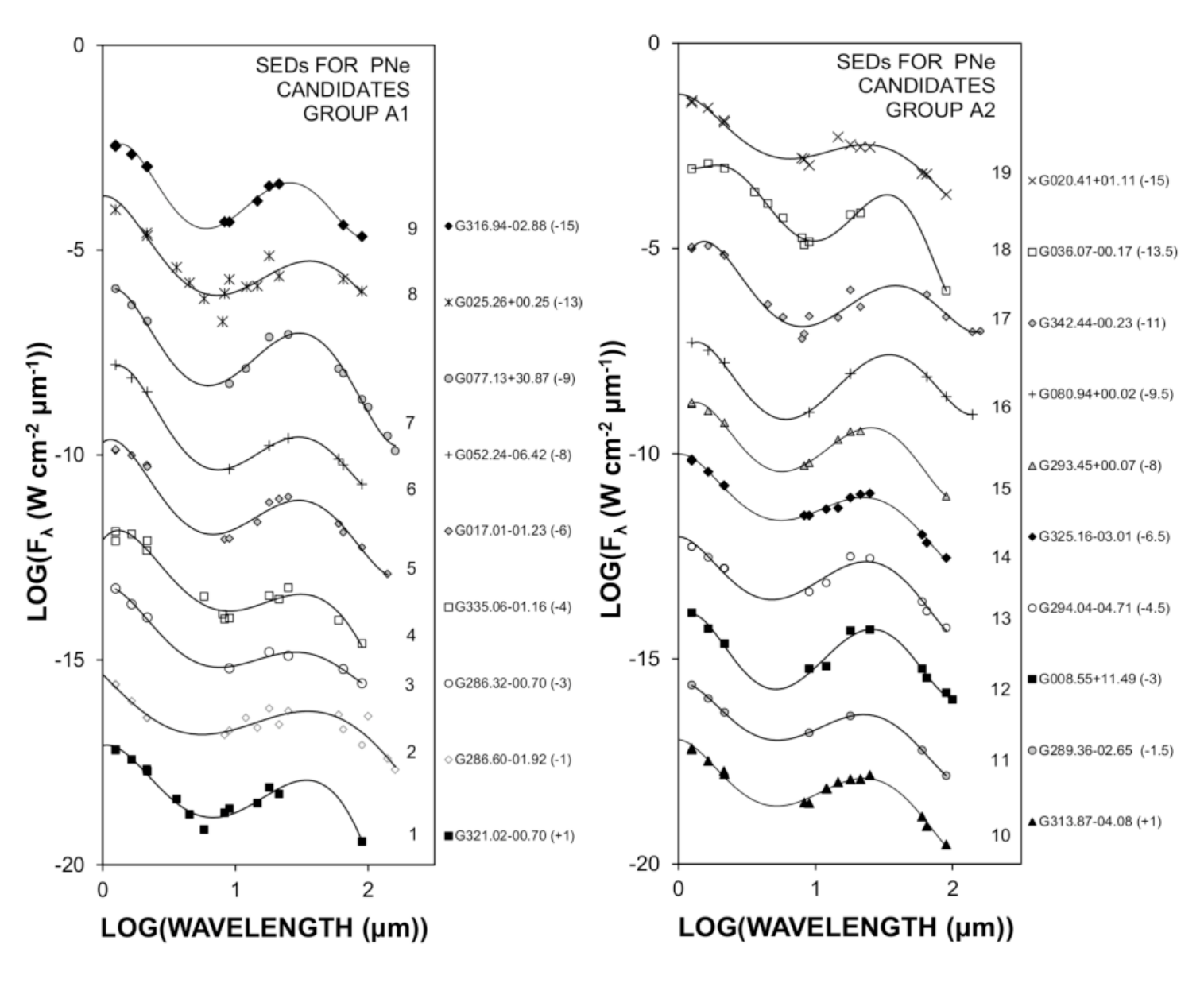}
\end{changemargin}
\caption{SEDs of 19 candidates in PNe group A, which show the double-component structure characteristic of PPNe with a dominance of the hot component.}
\label{PPN CandA1A2}
\end{figure}

\section{Spectral Energy Distributions}

It is anticipated that the 67 candidates identified as PN will exhibit a range of evolutionary states. We intend to explore these states through a comprehensive analysis of their SEDs. PNe in their formative stage, known as PPNe, possess an envelope that is still in the process of expansion. In the initial stages, the envelope exhibits a nearly uniform temperature, leading to emissions with a profile closely resembling that of a black body corresponding to a specific temperature. Furthermore, the central star of the Planetary Nebula (CSPN) emits with a black body profile corresponding to a distinct temperature. In this initial phase, the outcome is what \citet{2000oepn.book.....K} refers to as a two-component SED. Here, the cooler component aligns with the envelope expelled during the Asymptotic Giant Branch (AGB) phase, while the warmer component corresponds to the stellar photosphere. Following this, as the envelope expands, its temperature gradient will intensify. The overall emission will then be a cumulative effect of multiple layers, each at a distinct temperature, along with the emission from the star. The peak emission from the star will shift towards longer wavelengths (\citet{1988ApJ...331..832H}, \citet{1989ApJ...346..265H}).

\begin{figure}[H]
\begin{changemargin}{-1.5cm}{-1.0cm}
\centering
\includegraphics[scale=0.6, trim=60 0 0 0,clip]{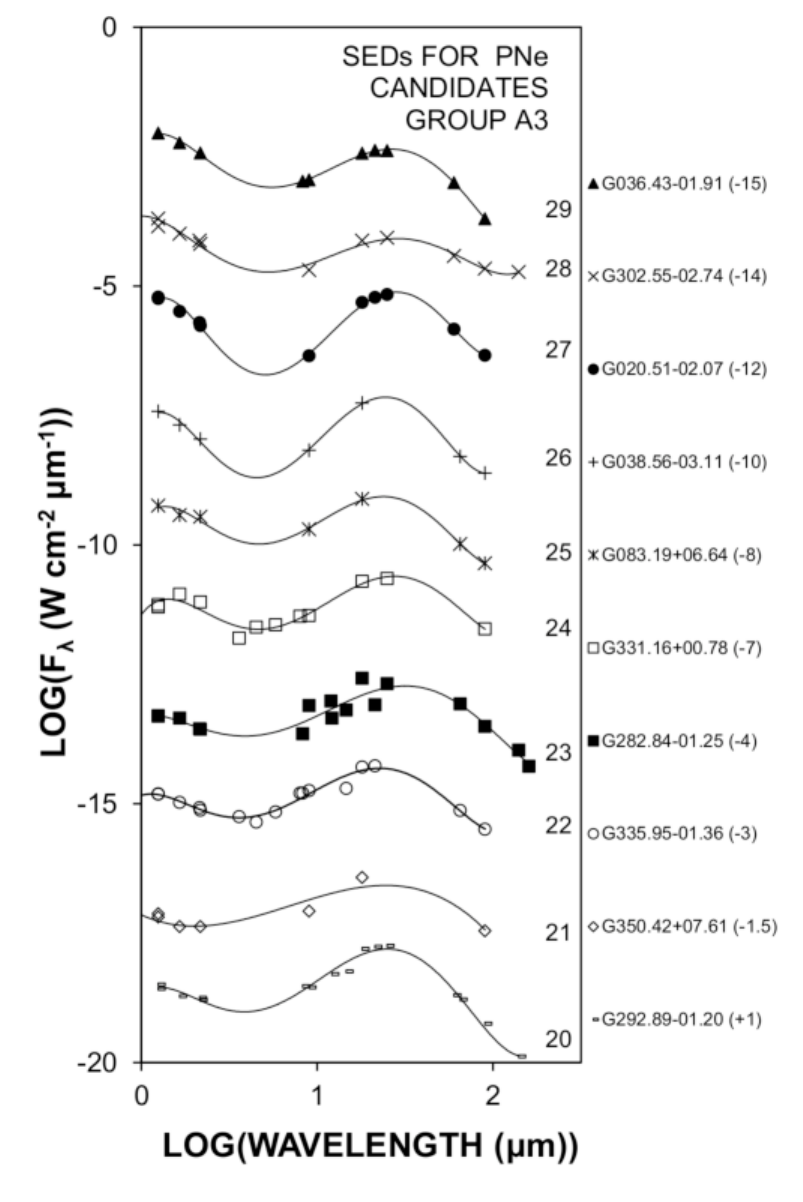}
\end{changemargin}
\caption{SEDs of 10 candidates in PNe group A, which show the double-component structure characteristic of PPNe with a relative balance between the cold and hot components.}
\label{PPN CandA3}
\end{figure}

To characterize the potential candidates, we utilized photometry across various observational projects, encompassing wavelengths from the blue, at 0.44$\mu$m (USNO B1 + B2), to the far-infrared (FIR) at 160$\mu$m (\emph{AKARI} FIS) (see Section Observations for details).  It is crucial to highlight that the photometric data from the USNO frequently produced outcomes that displayed disparities with those acquired from other telescopes. Consequently, this specific dataset was omitted from the construction of the SEDs for the associated sources.

\begin{figure}[H]
\begin{changemargin}{-1.5cm}{-1.0cm}
\centering
\includegraphics[scale=0.6, trim=60 0 0 0,clip]{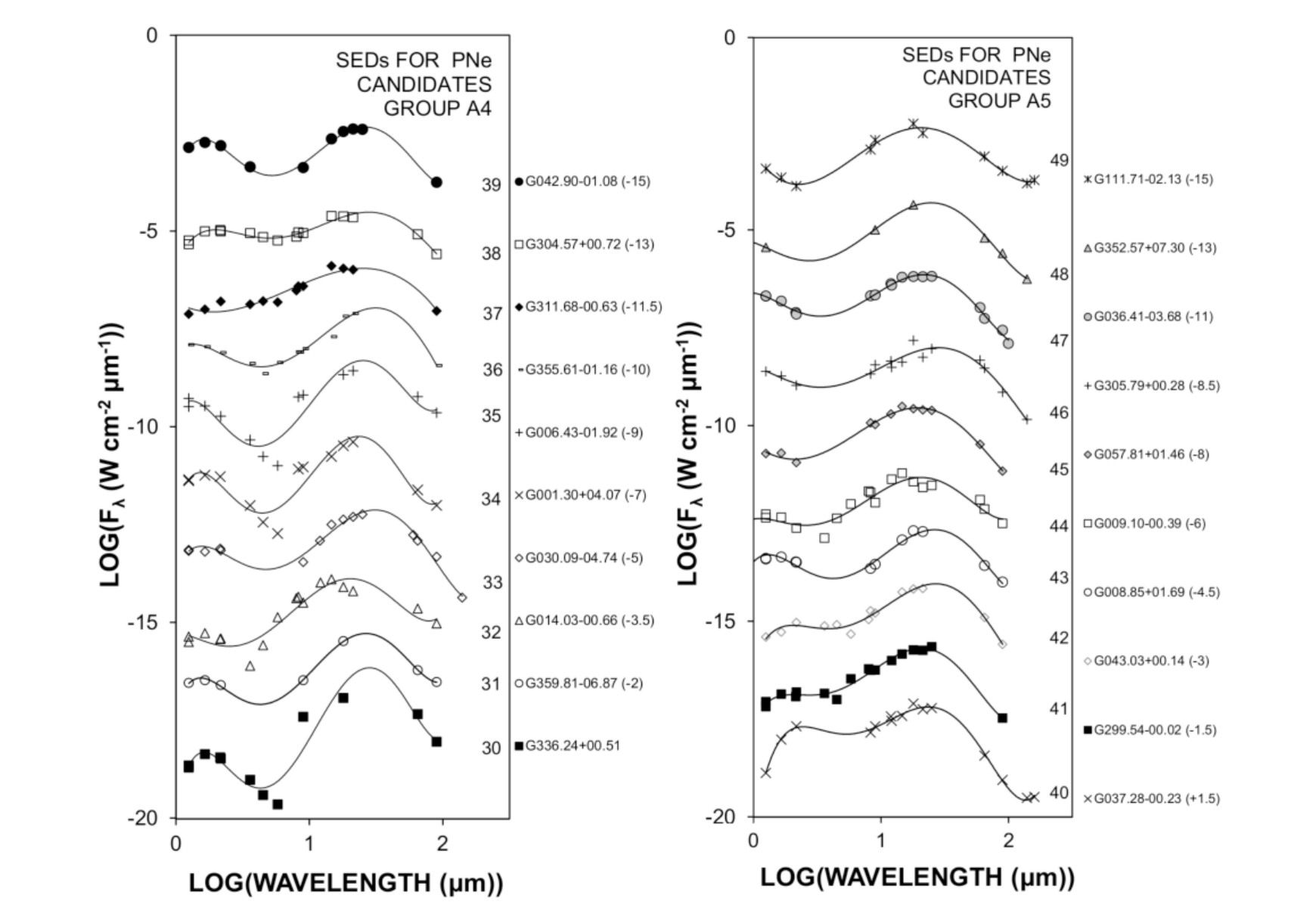}
\end{changemargin}
\caption{SEDs of 20 candidates in PNe group A, which show the double-component structure characteristic of PPNe with a dominance of the cold component.}
\label{PPN CandA4A5}
\end{figure} 

Furthermore, we produced comparable SEDs for well-established PNe and PPNe, serving as a control cohort. Figure \ref{Known PPN} illustrates the SEDs of confirmed PPNe, displaying their distinctive two-component architecture. Likewise, Figure \ref{PM PPN} showcases the SEDs of seven PPNe candidates identified by \citet{1988A&AS...76..317P}.

\begin{figure}[H]
\includegraphics[scale=0.5, trim=0 0 0 0,clip]{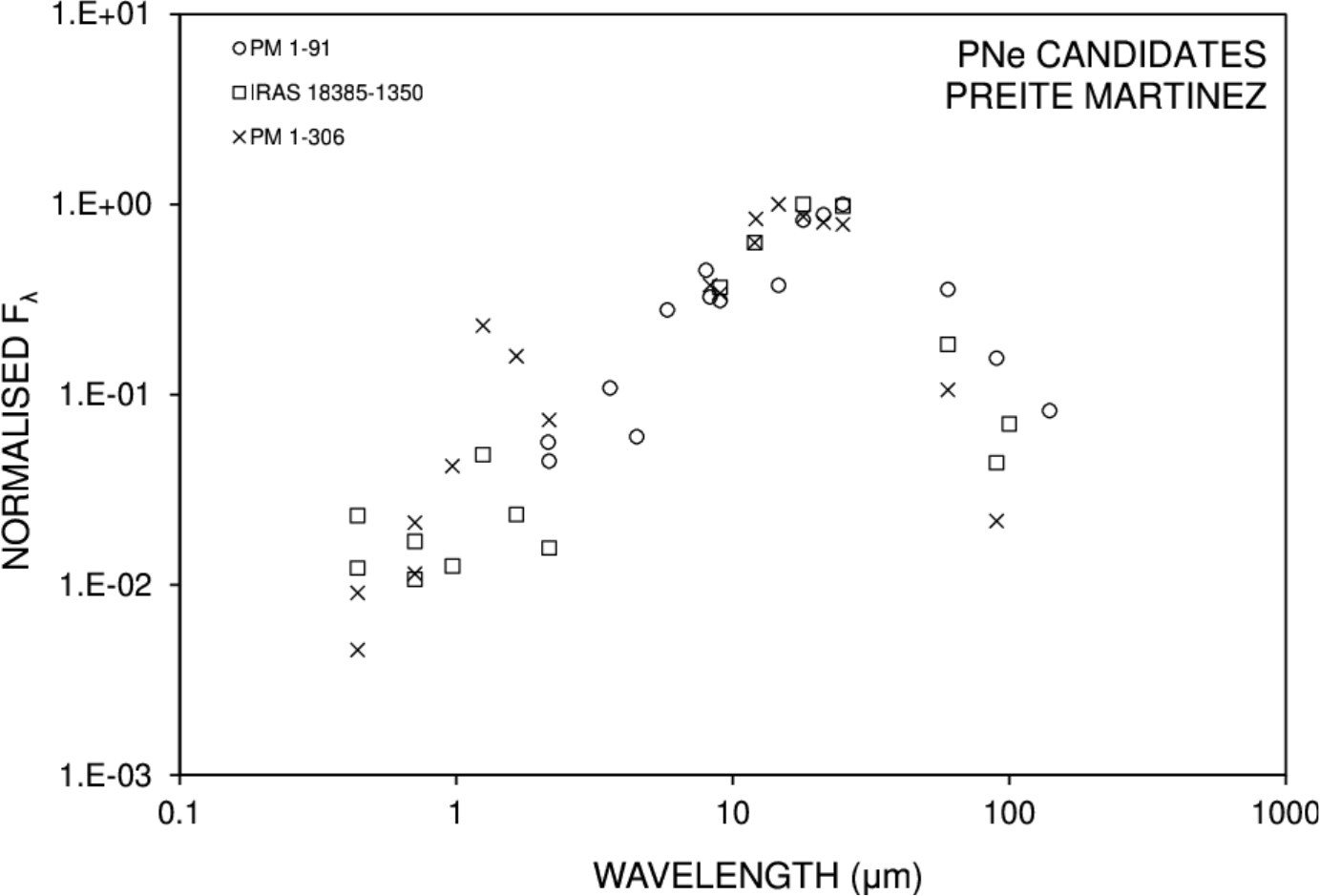}
\caption{SEDS of PN from \citet{1988A&AS...76..317P}. More evolved PNe which do not have a SED of two components but have layers with a continuous distribution of temperatures.}
\label{PM PN}
\end{figure}  

\begin{figure}[H]
\includegraphics[scale=0.43, trim=0 0 0 0,clip]{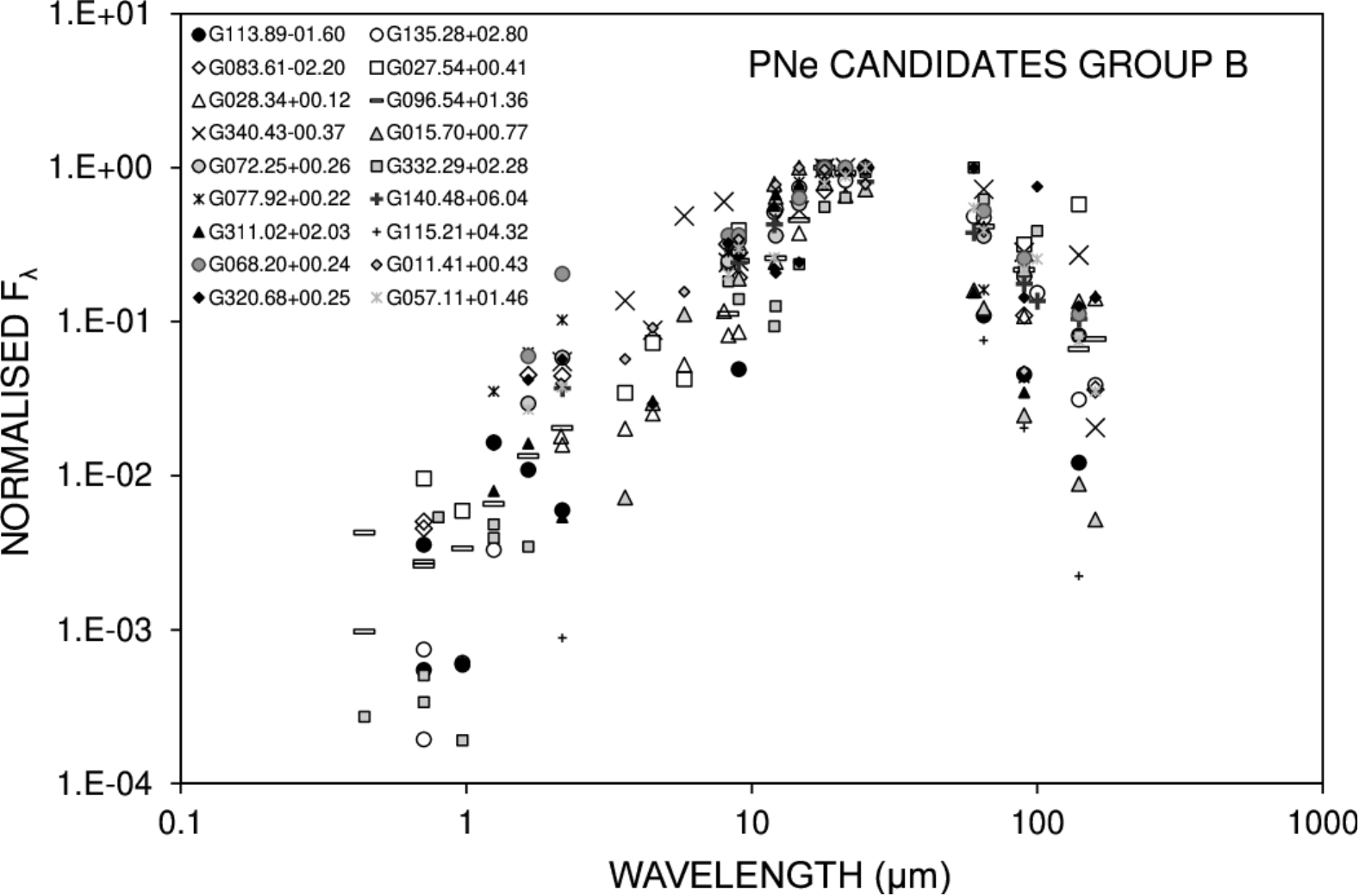}
\caption{SEDs of candidates in PNe group B, which have a more advanced evolutionary state.}
\label{PN Cand}
\end{figure}

Figures \ref{PPN CandA1A2}, \ref{PPN CandA3}, and \ref{PPN CandA4A5} illustrate the SEDs of 49 candidates displaying a two-component structure. We shall refer to this ensemble of 49 candidates, whose SEDs align with those of known PPNe, as Group A. For presentation purposes, we have subdivided Group A into five sets, labelled A1 to A5. In A1 and A2 (Figure \ref{PPN CandA1A2}), the SEDs are primarily influenced by the emission peak at shorter wavelengths, signifying the hotter component, i.e., the central star. In A3 (Figure \ref{PPN CandA3}), the SEDs display a relatively balanced emission from both components, while A4 and A5 (Figure \ref{PPN CandA4A5}) predominantly exhibit emissions from the cooler component, i.e., the envelope.

On the contrary, a distinct subset of candidates presents a discernible SED pattern, reminiscent of more advanced PNe that do not exhibit distinct components. Instead, they exhibit collective emissions from multiple layers, each distinguished by its individual black-body radiation. Figure \ref{PM PN} illustrates the SEDs of three objects previously identified by \citet{1988A&AS...76..317P} as PNe showcasing this SED profile. In Figure \ref{PN Cand}, we depict the SEDs of potential PN candidates characterised by an SED featuring only one peak. This subset of candidates, referred to as Group B, represents the more evolved candidates for PNe.

\section{Morphology and Profiles}
\label{mor}
In our initial endeavour to verify the nature of potential candidates, we examined the images from the \emph{AKARI} far-infrared all-sky survey maps. However, the pixel scale of the detectors, measuring 26.8$\arcsec$$\times$26.8$\arcsec$ for the short wavelength bands N60 and WIDE-S, and 44.2$\arcsec$$\times$44.2$\arcsec$ for the long wavelength bands N160 and WIDE-L \citep{2015PASJ...67...50D}, proved insufficient for drawing definitive conclusions. Consequently, we turned to the observations from \emph{WISE} (Wide-Field Infrared Survey Explorer, \citet{2010AJ....140.1868W}), to conduct an initial morphological study. This approach aids us in reviewing the list of potential candidates. 

 Figures \ref{WISE 1}, \ref{WISE 2} and \ref{WISE 3} present RGB images of the  67 potential candidates combining the W1  (in blue), W2 (green) and W3 (red) bands. Upon close examination of the RGB images, a noteworthy observation is that the majority of objects exhibit significant obscuration in the W1 and W2 bands in comparison to the W3 band. This characteristic is not observed in only two sources: G083.19+06.64 and G140.48+06.04 (Figure 10, Line 2, Column 3 and Line 4, Column 2 respectively). For the remaining sources, the prevalence of the W3 band is evident, ranging from subtle in the case of G077.13+30.87 to remarkably pronounced, as observed in G352.57+07.30. Another common feature among the majority of these sources is their radial symmetry, with the sole exception being G342.44-00.23. Two sources warrant additional commentary, G111.71-02.13 and G113.89-01.60 (Figure 10, Row 3, Columns 2 and 3 respectively). G111.71-02.13 appears to be associated with a larger arc-like structure and does not exhibit a star-like appearance. In Figure \ref{SN} a broader section of the sky encompassing this source is depicted (FoV 3.2'), with the highlighted area inside the white square corresponding to the segment presented in Figure 10. It is evident that G111.71-02.17 effectively constitutes a component of a Supernova Remnant (SNR), precisely identified as the SNR G111.7-02.1, commonly referred to as Cassiopeia A (Cas A). This enables us to exclude this source from the ultimate candidate list. On the other hand G113.89-01.60 seems to represent an emission peak within an arc encircling the star LS I +59 10, documented in the literature as a B0.2III spectral type star \citep{2003A&A...406..119N}. It is plausible to categorize G113.89-01.60 as a component of the HII region LBN 113.05-04.51, associated with LS I +59 10, and thereby, we can remove it from our candidate list too.

Furthermore, we have examined potential instances using \emph{Spitzer} IR images acquired with IRAC. In cases where IRAC images were unavailable, we turned to 2MASS images, aiming to identify morphologies consistent with extended emission of PNe, such as lobe structures, equatorial disks, or other PN-like configurations. Due to the limited resolution of 2MASS, we have opted to choose only certain cases of particular interest. Figure \ref{Some RGB} shows RGB images of nine sources, five from \emph{Spitzer} and four from 2MASS. For each instance, the orientation is set with North at the top and East to the left. The lower left corner of each image provides information about the data source, the lower right corner indicates the object name, and a scale bar is incorporated for reference.

 In Figure \ref{Some RGB}, five sources—G083.61–02.20, G113.89–01.60, G332.29+02.28, G027.54+00.41, and G135.28+02.80—exhibit a notably dense envelope that prevails over the central emission. Regarding this, the first two are the most prominent, where the radiation from a possible central star is totally absorbed by a spherical-type envelope; but G113.89 --01.60 appears in Figure \ref{WISE 2}, line 3, column 3 to be an extended part of another source, while G083.61 --02.20 presents an E-W structure. In the other three cases mentioned above, we observe a clear bipolar structure with progressively more emission in the centre, but this central emission does not dominate the lobe emission. These five sources are part of our group B of the general candidate list.

On the other hand, the images of G304.57 +00.72, G311.68 --0.63, G025.26 +00.25 and G321.02 --00.70 show dominant central stars. In the first two cases we see spherical envelopes absorbing the emission of the central star. In both images we observe the envelope heated by the stellar radiation, although, as we can see in their respective SEDs (Fig. \ref{PPN CandA4A5}, group A4, 38 and 37), the stellar emission is still important and its emission peak can be distinguished. Meanwhile, the image of G025.26 +00.25, although presenting an envelope similar to G304.57 +00.72 and G311.68 --00.63, has subtle differences which have important effects in its emission. The star is not situated in the centre of the envelope, which allows part of the stellar emission to escape. In the corresponding SED (Fig. \ref{PPN CandA1A2}, group A1, number 8) the dominant component is the high-temperature one (central star). Finally, the image of G321.02 --00.70 shows a structure clearly reminiscent of bipolar PNe. The emission from the central star dominates the combined emission from both lateral structures located to the NE and SW, as can be seen in its SED (Fig. \ref{PPN CandA1A2}, A1, 1), which shows that the hot component clearly dominates the cold one.

\begin{figure}[H]
\begin{changemargin}{0cm}{0cm}
\includegraphics[scale=0.6, trim=0 0 0 0]{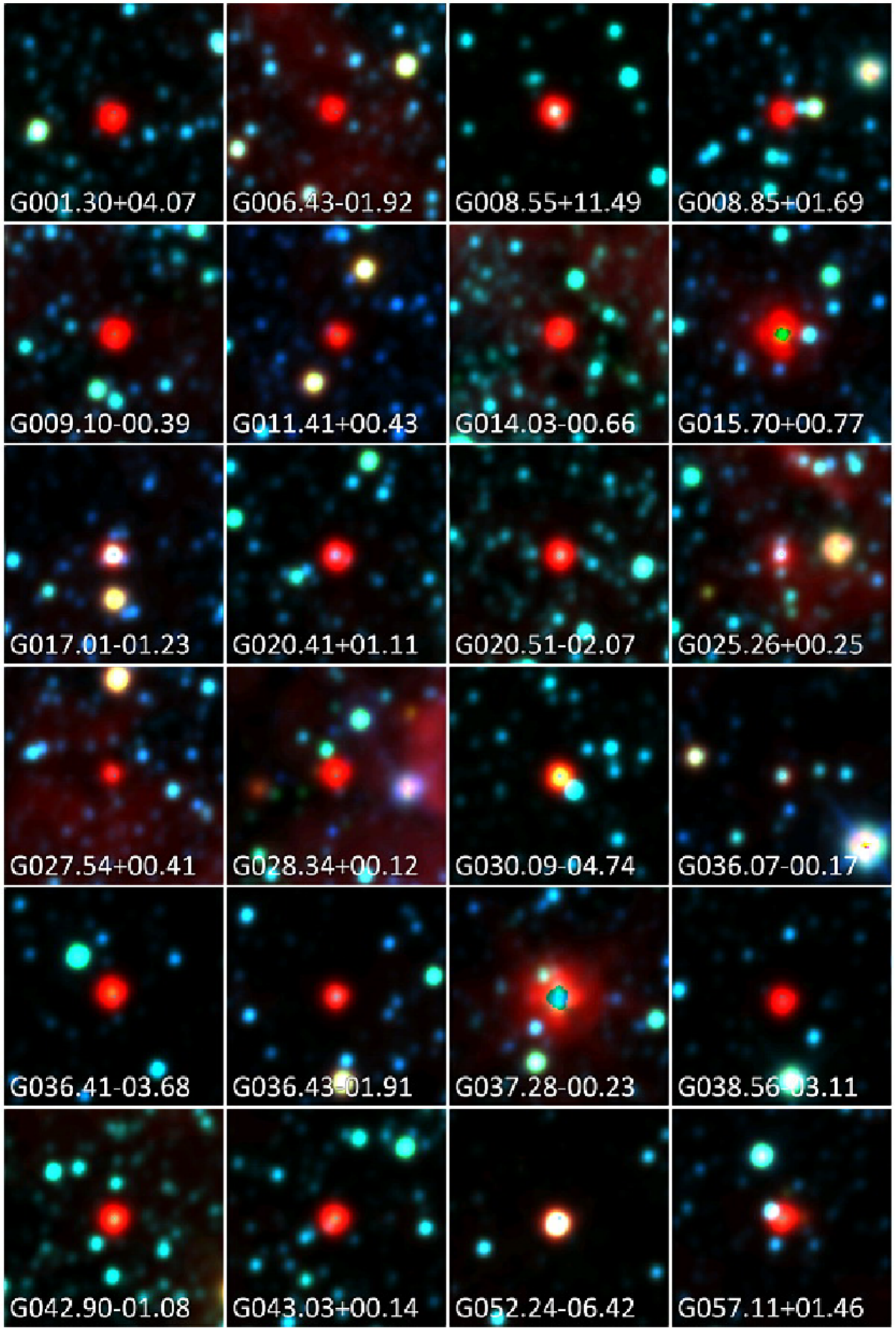}
\end{changemargin}
\caption{RGB images of PNe candidates 1 to 24. RGBs constructed with \emph{WISE} bands 3.4$\mu$m (blue), 4.6$\mu$m (green) and 12$\mu$m (red). In all images north is up, east to the left, and FoV 3.2'.}
\label{WISE 1}
\end{figure} 

\begin{figure}[H]
\begin{changemargin}{0cm}{0cm}
\includegraphics[scale=0.6, trim=0 0 0 0]{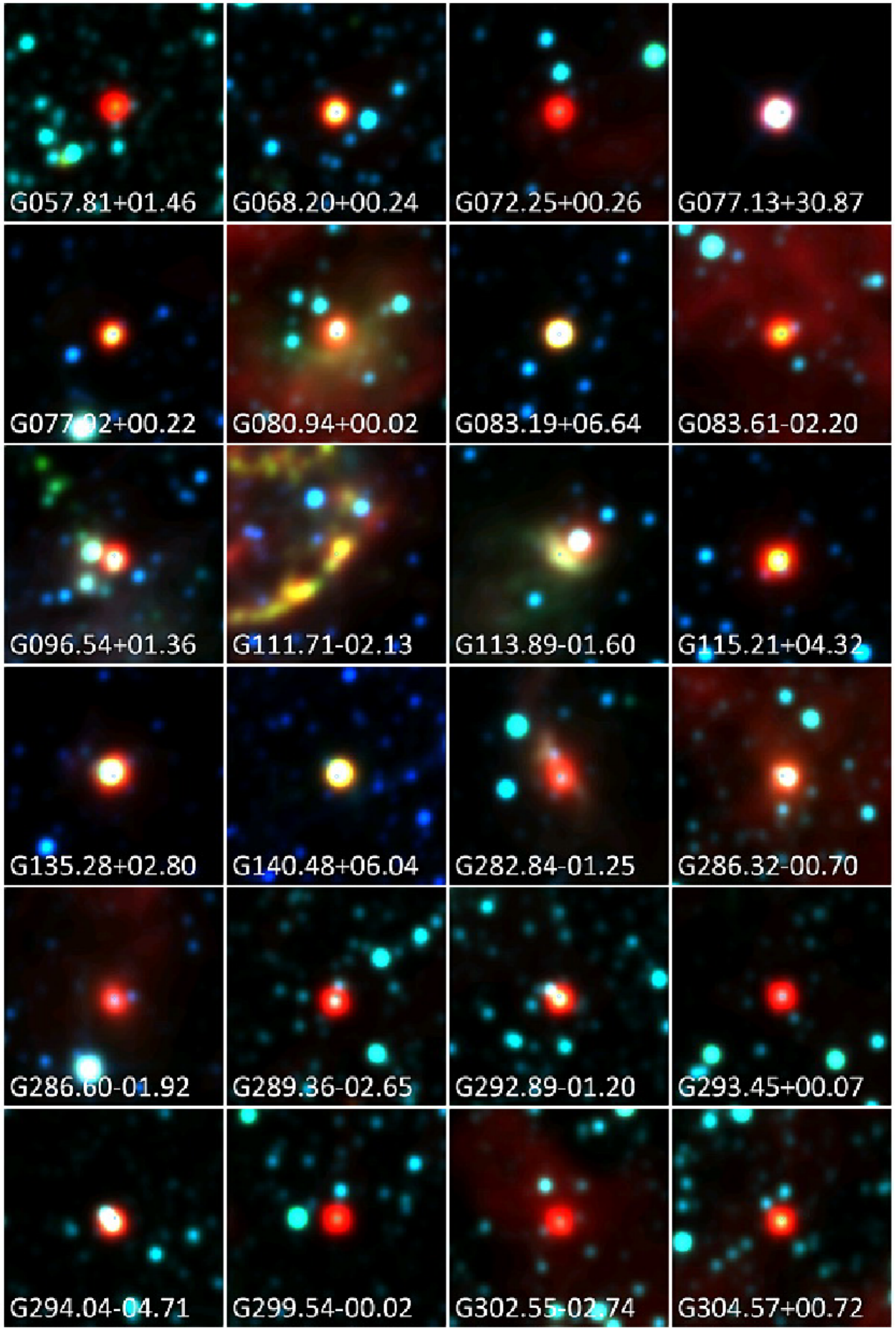}
\end{changemargin}
\caption{RGB images of PNe candidates 25 to 48. RGBs constructed with \emph{WISE} bands 3.4$\mu$m (blue), 4.6$\mu$m (green) and 12$\mu$m (red). In all images north is up, east to the  left, and FoV 3.2'.}
\label{WISE 2}
\end{figure} 

\begin{figure}[H]
\begin{changemargin}{0cm}{0cm}
\includegraphics[scale=0.6, trim=0 0 0 0]{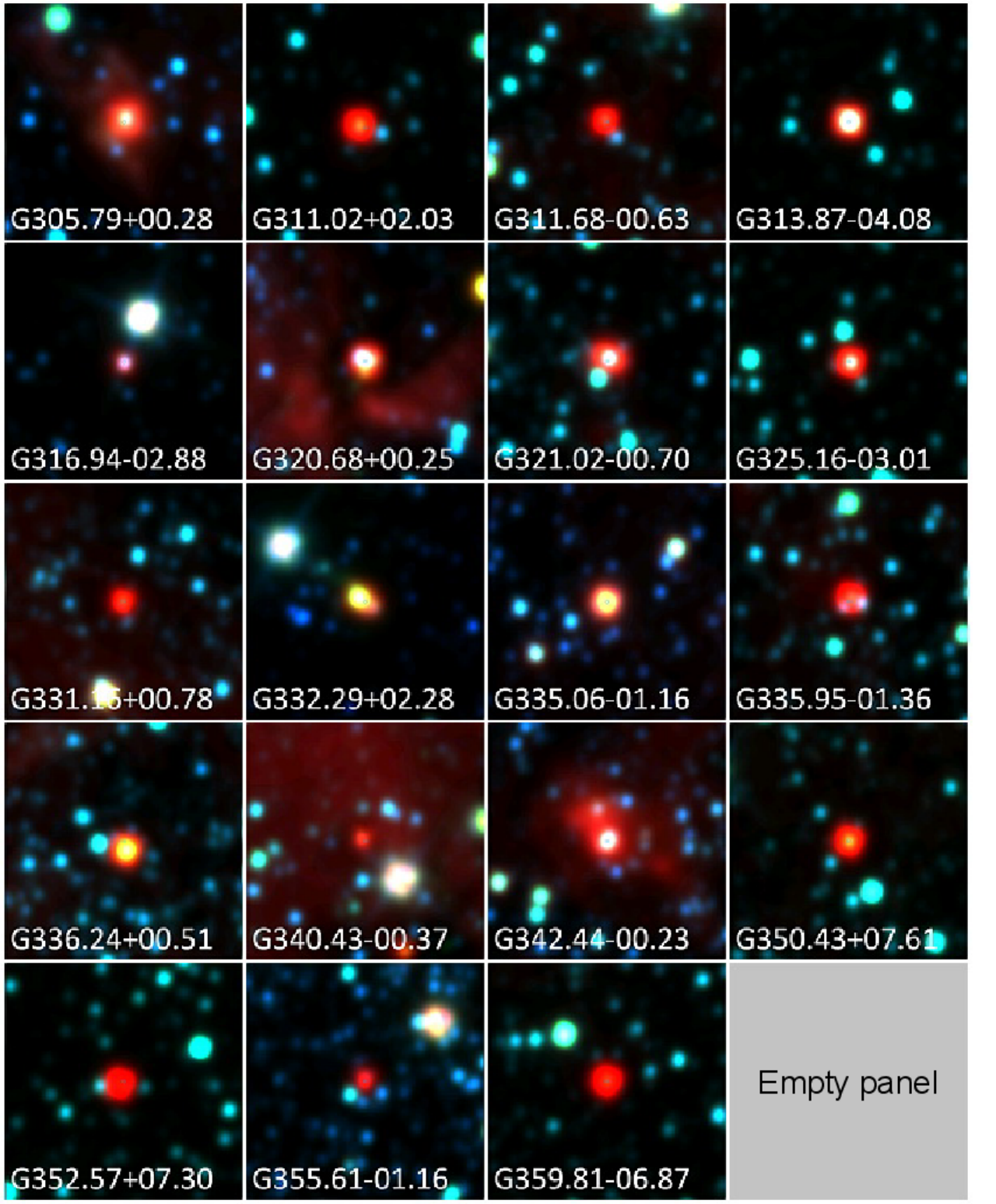}
\end{changemargin}
\caption{RGB images of PNe candidates 49 to 67. RGBs constructed with \emph{WISE} bands 3.4$\mu$m (blue), 4.6$\mu$m (green) and 12$\mu$m (red). In all images north is up, east to the left, and FoV 3.2'.}
\label{WISE 3}
\end{figure}

\begin{figure}[H]
\begin{changemargin}{2.5cm}{0cm}
\includegraphics[scale=0.25, trim=0 0 0 0, clip]{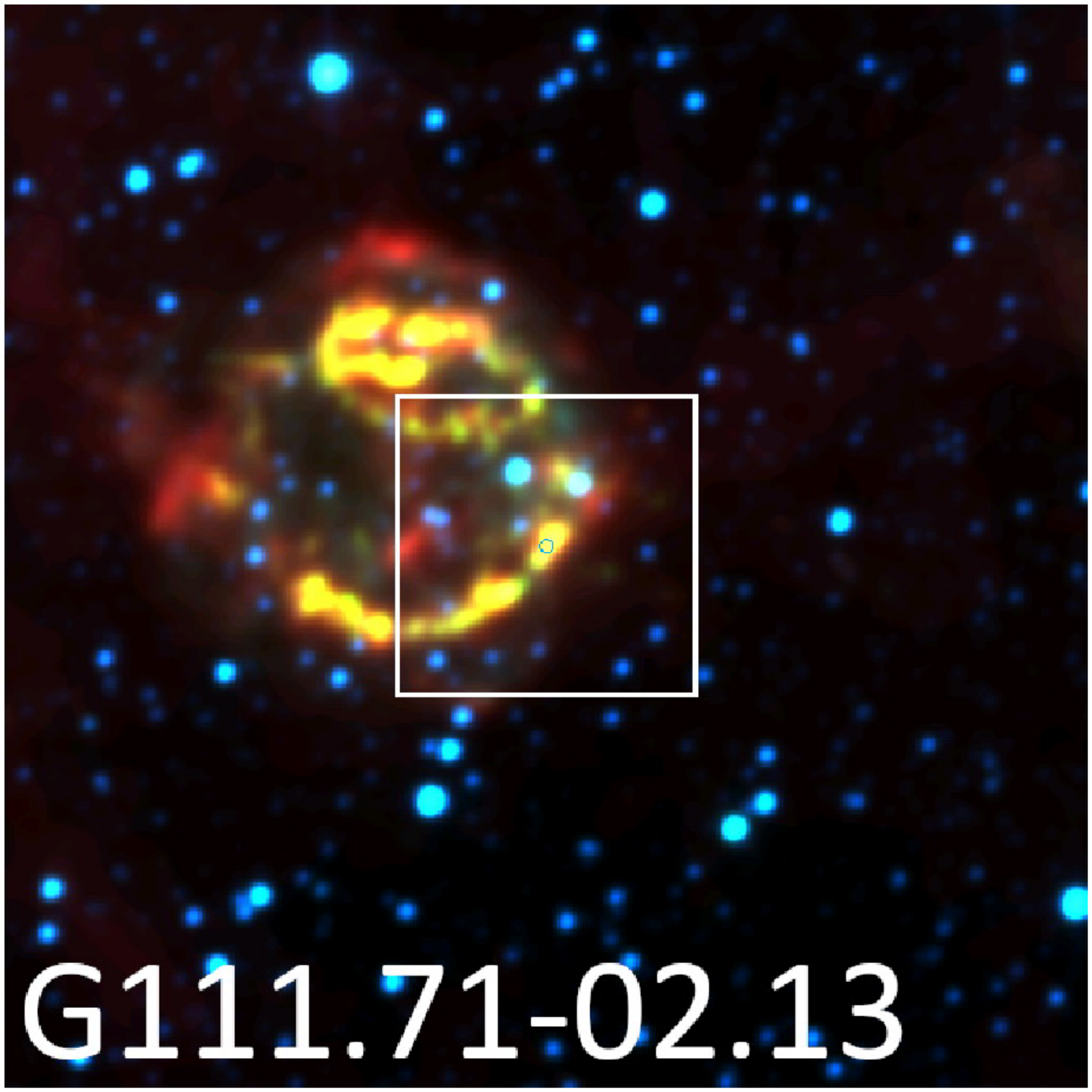}
\end{changemargin}
\caption{Wide RGB images of PNe candidate G111.71-02.13. RGBs constructed with \emph{WISE} bands 3.4$\mu$m (blue), 4.6$\mu$m (green) and 12$\mu$m (red). North is up, east to the left, and the total FoV is 12' while the white box encloses the frame shown in Figure \ref{WISE 2} with FoV of 3.2'. }
\label{SN}
\end{figure}

\begin{figure}[H]
\begin{changemargin}{0.5cm}{0cm}
\includegraphics[scale=0.35, trim=0 0 0 0]{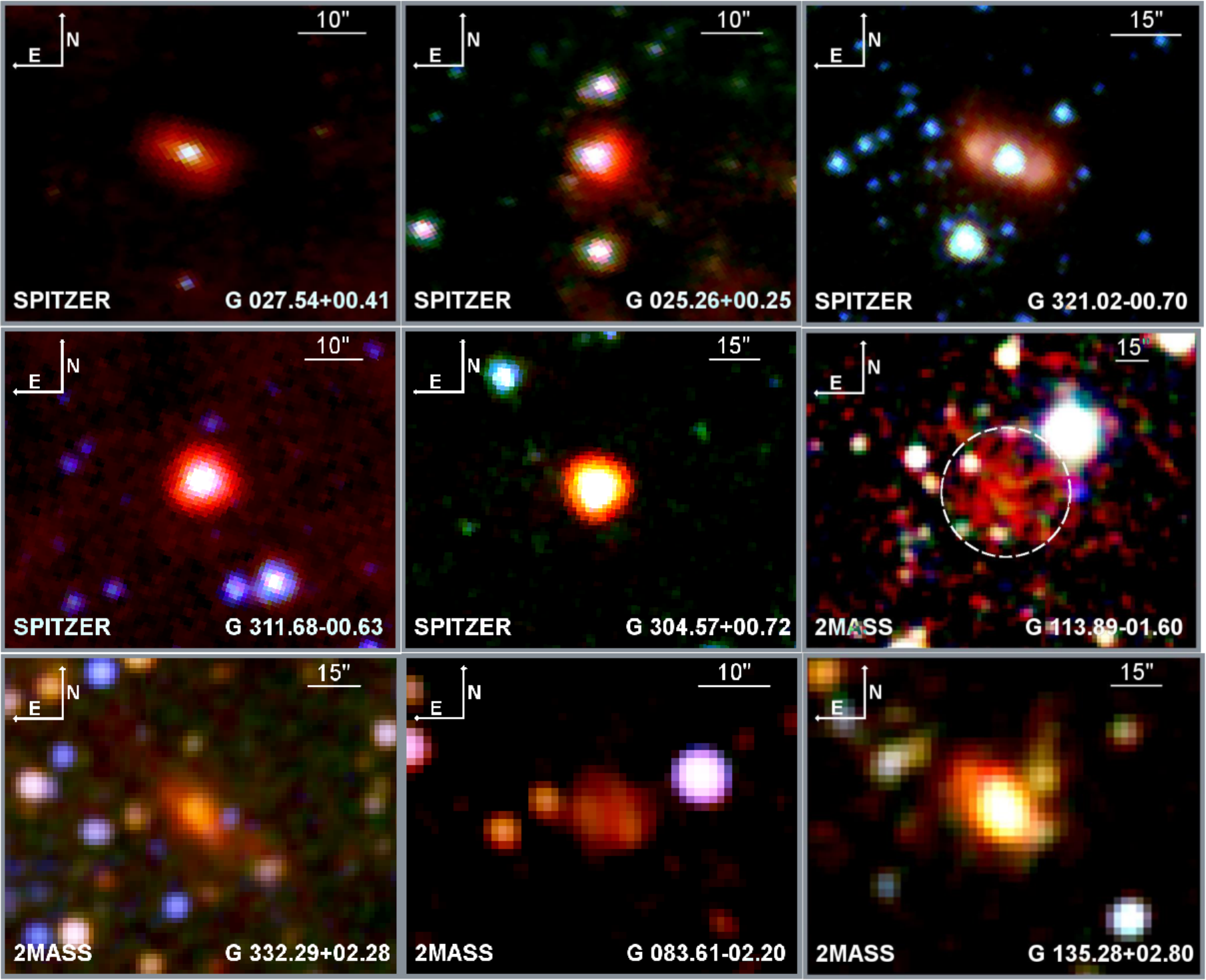}
\end{changemargin}
\caption{RGB image of PNe candidates. The RGBs constructed from \emph{Spitzer} images use the 4.5$\mu$m (blue), 5.8$\mu$m (green) and 8.0$\mu$m (red) bands, while those from 2MASS use the 1.25$\mu$m (blue), 1.65$\mu$m (green), and 2.17$\mu$m (red) bands.}
\label{Some RGB}
\end{figure} 

\begin{figure}[H]
\begin{changemargin}{-1.0cm}{0cm}
\includegraphics[scale=0.65, trim=0 0 0 0,clip]{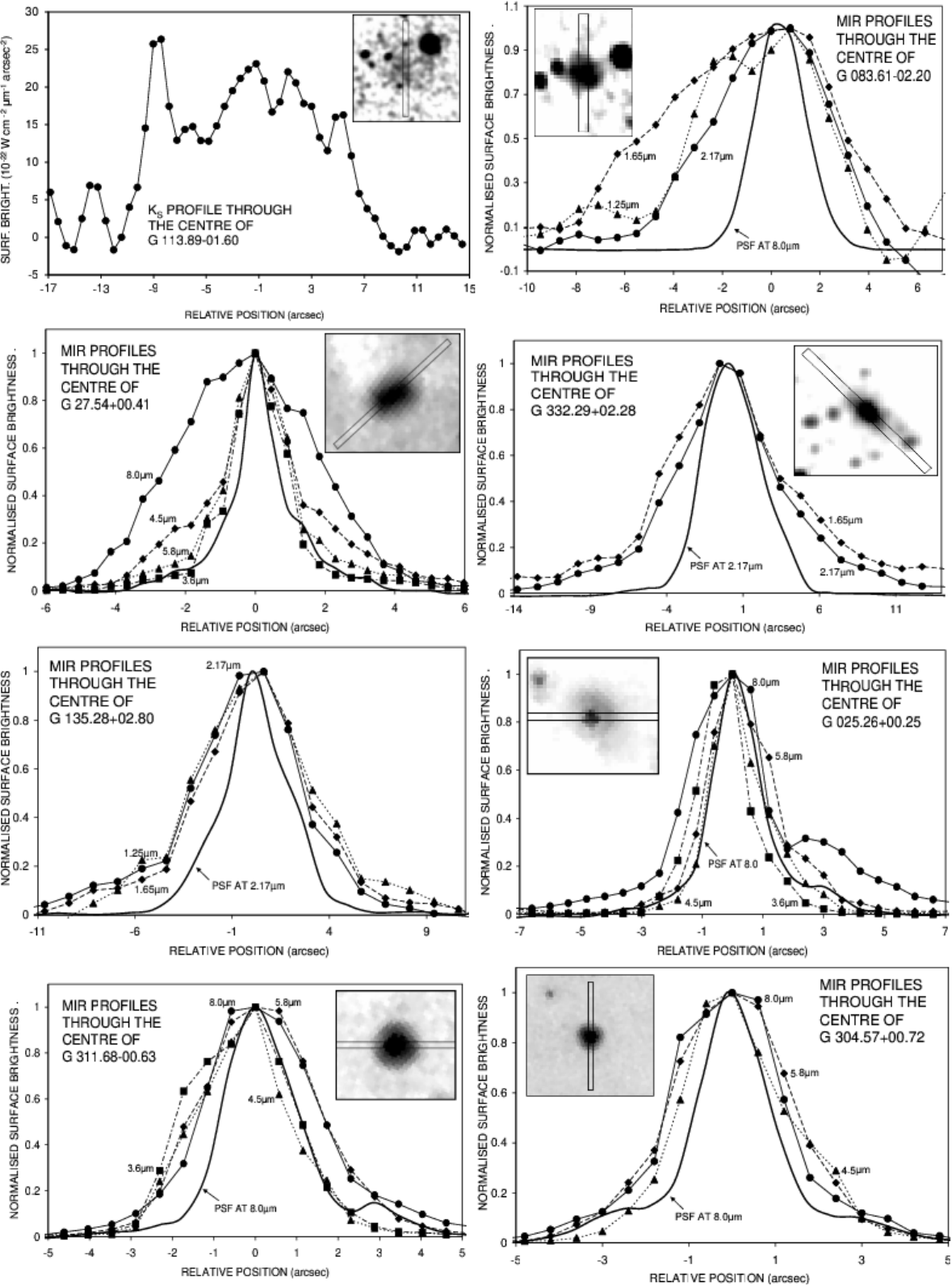}
\end{changemargin}
\caption{Spatial profiles of selected candidates. We have extracted spatial profiles along an axis through the centre of a sample of candidates, in order to study the spatial distribution of the emission and determine if they are extended sources.}
\label{Pofiles}
\end{figure} 

We have conducted a comprehensive examination of the spatial distribution of the emission across multiple available bands for a subset of the candidates. The extracted profiles are depicted in Figure \ref{Pofiles}. The profiles of G113.89 --01.60 and G083.61 --02.20 show that these are extended sources. In particular the profile of G113.89 --01.60 has K-band emission with a width of $\sim 22$ arcsec, but this emission appears in Figure \ref{WISE 2}, line 3, column 3 as an extended part of another source. G083.61 --02.20 has a width of $\sim 10$ arcsec in the three 2MASS bands. G027.54 +00.41 and G332.29 +02.28 show an elongated emission profile, which is expected for PNe with a certain degree of evolution, and this seems to be supported in the SEDs of both objects in Figure 8. In both cases the images also show the elongated structure characteristic of bipolar PNe.

G135.28 +02.80 and G025.26 +00.25 also show extended profiles, although with convex-shaped sides which appear to be layers expelled from the central region. In G135.28 +02.80 the convex-shaped sides are quite symmetric, which suggests that the central area is embedded in a dusty envelope, while in the case of G025.26 +00.25 this envelope is clearly asymmetric, permitting the emission of the possible CSPN to escape freely, which is confirmed by its SED (Fig. \ref{PPN CandA1A2}, A1, 8). Finally the profiles of G311.68 --00.63 and G304.57 +00.72 show slightly extended emission in the four IRAC bands, especially at 5.8$\mu$m and 8.0$\mu$m. This can be explained by a layer of expelled dust, which can be confirmed using their SEDs (Fig. \ref{PPN CandA4A5}, A4, 37 and 38 respectively), where two similar-strength components of thermal emission can be seen.

\section{Independent Confirmation of G050.48 --00-70}

One of the sources within the PN region (Fig. \ref{Colour-Colour Diagram}) that has been excluded from the current list of candidates is G050.48 –00.70. This object was previously identified as a PN by \citet{2009A&A...507..795U}, in a study on YSOs that is entirely independent of the analysis conducted here. To verify this classification, we present images (Figure \ref{G050.48}) and emission profiles (Figure \ref{Profile color}) of G050.48 --00.70 from \emph{Spitzer}/IRAC, 2MASS and IPHAS. The H$\alpha$ emission reveals significant extinction in its central region (Fig. \ref{G050.48}, above), with a pair of symmetric `cones' of emission escaping from the high extinction region.  This suggests the presence of dusty material in the centre and ionized material in the cones. The color-composite image with J (blue), H (green) and K (red) from 2MASS (Fig. \ref{G050.48}, centre) exhibits strong emission in the center of the cones and between them at shorter wavelengths, while K-band emission dominates at the edges. This indicates that the conical structures are being heated from the central region. Finally the color-composite image with bands 4.5$\mu$m (blue), 5.8$\mu$m (green) and 8.0$\mu$m (red) from \emph{Spitzer}/IRAC (Fig. \ref{G050.48}, below) illustrates emission concentrated in the central region across all three bands.

\begin{figure}[H]
\begin{changemargin}{2.5cm}{1cm}
\includegraphics[scale=0.36, trim=0 0 0 0,clip]{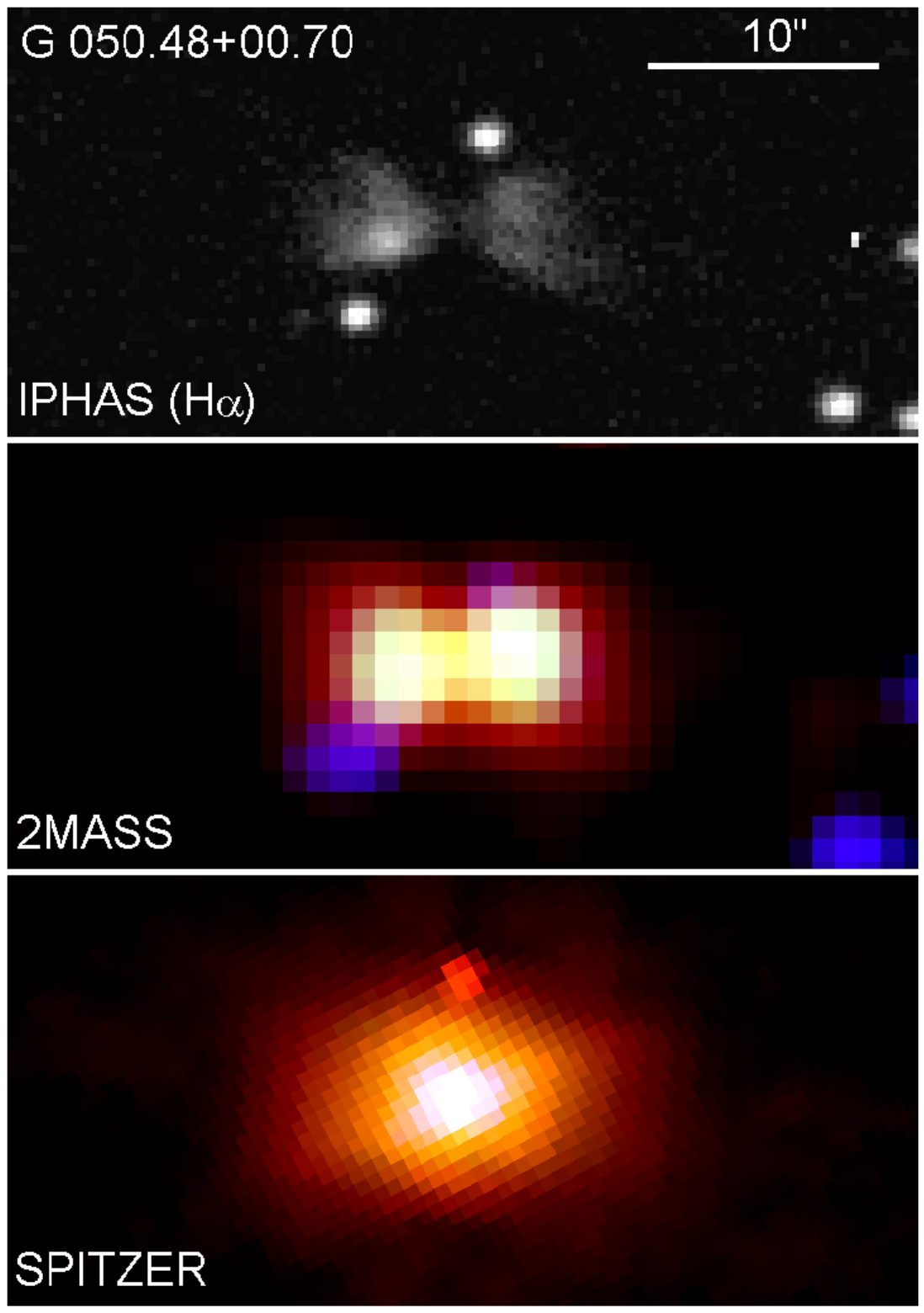}
\end{changemargin}
\caption{Images of G050.48 +00.70. Above: H$\alpha$ $\lambda$656.3nm from IPHAS. Centre: Composite color image with J (blue), H (green), K (red) from 2MASS. Below: Composite color image using the  4.5$\mu$m (blue), 5.8$\mu$m (green) and 8.0$\mu$m (red) bands from \emph{Spitzer}/IRAC.}
\label{G050.48}
\end{figure} 

\begin{figure}[H]
\includegraphics[scale=0.40, trim=0 0 0 0,clip]{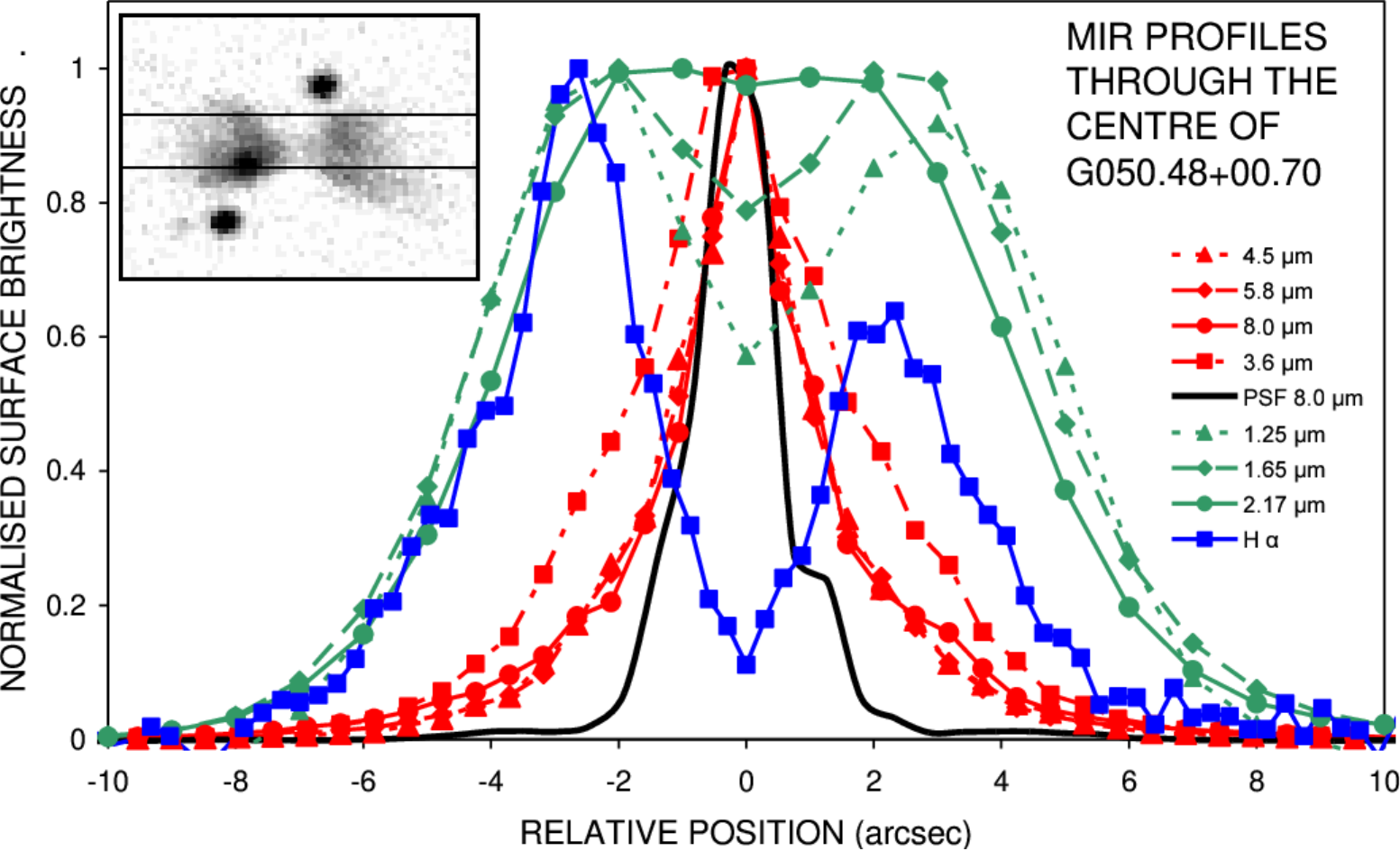}
\caption{Spatial profiles of G050.48 +00.70. The H$\alpha$ emission profile is in blue, the three 2MASS bands in green and the four \emph{Spitzer}/IRAC bands in red. Each 2MASS and \emph{Spitzer}/IRAC band is identified by different geometric shapes.}
\label{Profile color}
\end{figure} 

In particular, the emission at 8.0$\mu$m indicates heated dust, which is the same material that causes the extinction in H$\alpha$, then re-emitting at 8.0$\mu$m. All the above can be seen in the graphical representation of the emission profiles in Fig. \ref{Profile color}, which contains the H$\alpha$ profiles, the 3 bands of 2MASS and the four from \emph{Spitzer}/IRAC.  Therefore, from the PSC photometry of \emph{AKARI}, and the morphology and emission profiles in the optical, NIR and FIR, we can conclude independently of \citet{2009A&A...507..795U} that G050.48 +00.70 is a PN.

\section{Conclusions}

By analyzing the \emph{AKARI} photometry and other instruments employed in this study, we can draw the following conclusions.

i) The photometry in the 9$\mu$m, 18$\mu$m y 90$\mu$m  \emph{AKARI} bands enables the determination of colour indices that are effective in discriminating between PNe-type sources, and other astronomical objects.

ii) At least 67 sources from the \emph{AKARI} PSC exhibit emission characteristics consistent with PNe or related objects, yet they haven't been definitively identified as such in the existing literature. We consider these as potential candidate PNe and/or PPNe.

iii) Examination of the SEDs of the potential candidates facilitates their categorization into two groups based on their evolutionary stage, PPNe and PNe.

iv) Visual inspection of images allows us to exclude two sources, G113.89-01.60, identified as part of a stellar envelope, and G111.71-02.13, associated with the Cassiopeia A SNR.

v) Finally, after the elimination of two potential candidates, we propose 65 sources from the \emph{AKARI} PSC as PNe or PPNe candidate.

\hspace{1cm}

Additionally, we propose two avenues for future research. Firstly, a spectroscopic examination of the 65 final candidates to either validate or discard their candidacy. Secondly, a reconsideration of the limits used to define the Planetary Nebulae (PNe) region in our colour-colour diagram. If the majority of the 65 candidates presented here are substantiated, it would be valuable to expand the boundaries of the PN region, encompass more sources, categorize them, and subject them to thorough analysis of images and spectra for confirmation or rejection as candidates.

\section*{Acknowledgments}
Authors express their gratitude to the Mexican National Council for Science and Technology (CONACyT) for financing this work.
GR-L acknowledges support from CONACYT (grant 263373) and PRODEP (Mexico).
This research has made use of the NASA/IPAC Infrared Science Archive, which is operated by the Jet Propulsion Laboratory, California Institute of Technology, under contract with the National Aeronautics and Space Administration. This research has made use of the SIMBAD database, operated at CDS, Strasbourg, France. This research is based on observations with \emph{AKARI}, a JAXA project with the participation of ESA. This work is based on observations made with the INT and WHT telescopes operated on the island of La Palma by the Isaac Newton Group (ING) in the Spanish Observatorio del Roque de los Muchachos.

\end{document}